\documentclass[oneside,letter,12pt,pdflatex]{article}
\usepackage[utf8]{inputenc}
\usepackage[margin=1.25in]{geometry}
\usepackage{natbib}
\usepackage{booktabs,calc}
\usepackage{amssymb}
\usepackage{amsmath}
\usepackage{bbm}
\usepackage{amsthm} 
\usepackage{threeparttable}
\usepackage{epsfig}
\usepackage{graphics}
\usepackage{pdflscape}
\usepackage{multirow}
\usepackage{tikz}
\usepackage{comment}
\usepackage{subcaption}
\usepackage{algorithm2e}
\usepackage{array}
\usepackage{rotfloat}
\usepackage{units}
\usepackage{textcomp}
\usepackage{mathtools}
\usepackage{graphicx}
\usepackage{geometry}
\usepackage{pdflscape}
\usepackage{siunitx}
\usepackage{soul}
\usepackage{xcolor}
\usepackage[nodisplayskipstretch]{setspace}

\PassOptionsToPackage{normalem}{ulem}
\usepackage{ulem}
\usepackage{microtype}

\newcommand{\bi}{\begin{itemize}}
\newcommand{\ei}{\end{itemize}}
\newcommand{\beq}{\begin{equation}}
\newcommand{\eeq}{\end{equation}}
\newcommand{\be}{\begin{equation}}
\newcommand{\ee}{\end{equation}}

\definecolor{winered}{rgb}{0.5,0,0}

\usepackage[colorlinks=true, linkcolor=winered, citecolor=winered, urlcolor=black, bookmarks=true]{hyperref} 
\usepackage{bookmark}

\begin{document}
 \begin{spacing}{1}

\title{(Machine) Learning What Policies Value\thanks{
We thank Luk Yean and Jolie Wei for excellent research assistance. Thank you to Joseph Cummins, Brian Dillon, John Friedman, Ted Miguel, Sendhil Mullainathan, Paul Niehaus, Yang Xie, and seminar audiences for helpful conversations. We thank the JP Morgan Chase Research Assistant Program at Brown University for financial support.}}
\author{
		Daniel Bj{\"o}rkegren\footnote{\small dan@bjorkegren.com} \\ 
			\footnotesize Brown University \ 
		\and
		Joshua E. Blumenstock\footnote{\small jblumenstock@berkeley.edu} \\ 
			\footnotesize U.C. Berkeley \ 
		\and
		Samsun Knight\footnote{\small samsun\_knight@brown.edu}\\ 
	  		\footnotesize Brown University \
	}

% \date{\small This version: \today}

\maketitle
\thispagestyle{empty}

\begin{abstract}
% \noindent \small 
When a policy prioritizes one person over another, is it because they benefit more, or because they are preferred?
This paper develops a method to uncover the values consistent with observed allocation decisions. We use machine learning methods to estimate how much each individual benefits from an intervention, and then reconcile its allocation with (i) the welfare weights assigned to different people; (ii) heterogeneous treatment effects of the intervention; and (iii) weights on different outcomes. We demonstrate this approach by analyzing Mexico's PROGRESA anti-poverty program. The analysis reveals that while the program prioritized certain subgroups --- such as indigenous households --- the fact that those groups benefited more implies that they were in fact assigned a lower welfare weight. The PROGRESA case illustrates how the method makes it possible to audit existing policies, and to design future policies that better align with values.

%This paper develops a method to uncover the values consistent with observed allocation decisions. We use machine learning estimators for heterogeneous treatment effects to identify who benefits from an allocation. We then decompose the objective underlying the allocation into: differential (i) treatment effects, (ii) welfare weights between entities; and (iii) impact weights across outcomes. We apply this approach to Mexico's Progresa anti-poverty program, to estimate the preferences its allocation implies. The impacts of the program were heterogeneous, especially along income and age of household head. Controlling for these differential impacts, allocations are consistent with welfare weights that rank a household 13 percentiles higher if indigenous, 8 percentiles lower for each standard deviation increase in household income, and 21 percentiles higher for each additional small child in the household, on average. Allocations are consistent with valuing each missed school day and child sick day within conventional valuations, though our estimates are imprecise. Alternate eligibility criteria could have improved either average consumption, health or schooling outcomes.
\end{abstract} 

\vspace{1cm}


\vspace{.5cm}
\vspace{2cm}

\end{spacing}

\pagebreak

\section{Introduction}

The values behind policy decisions are not always transparent. When governments decide which households receive welfare benefits, or universities select which students to admit, they do not always articulate the rationale behind those decisions. Even when policymakers do articulate a rationale, it may be difficult to verify.
%Governments decide which households receive welfare benefits, but do not always articulate the rationale behind eligibility criteria. Universities select students to admit, but the preferences behind those decisions are often opaque. 
In particular, certain people may be prioritized either because they are expected to benefit the most from the policy, or because they are favored --- irrespective of how much they are likely to benefit. 
%Allocation decisions like these may prioritize certain entities because those entities benefit most, or because they are intrinsically valued, separately from how much they benefit. 
This distinction has important implications \citep{nichols_targeting_1982,coate_form_1995}: all members of society may agree on a ranking of who benefits most along some objective metric, but may disagree on how much welfare weight to assign to different entities.

This paper develops a method to infer social preferences that are consistent with observed policies. This method relies on recent developments in machine learning, which make it possible to estimate differential treatment effects without overfitting. We show how such methods can be combined with a second stage regression to separate heterogeneous treatment effects (who benefits the most) from implied welfare weights (who is valued) and how different outcomes are valued. As a result, we can shift the debate from one about means --- who should receive what --- to one about ends --- what are the impacts we desire, and which populations are most important?

%We demonstrate how these advances can be used to better understand and articulate the allocation of programs. Our approach allows us to pose counterfactual welfare weights and valuations of outcomes to produce different allocations, and quantify the welfare impacts of these adjustments.

We consider a common form of decision, an allocation based on a score or ranking. These may be poverty scores in the case of welfare programs,
or explicit rankings in the case of applicants for college admission or small business grants. This ranking implies a system of inequalities between the contributions of different entities to welfare. We use this system of inequalities, and a simple and general model of welfare, to estimate the implied value on different outcomes (estimated using modern methods for heterogeneous effects) and different entities (based on observed characteristics), using ordinal logistic regression. Our method can also be used if one only observes the binary decision of who is eligible and who is not, though it will have less power.

Intuitively, if a policy allocates benefits to one type of entity who is expected to benefit little from the allocation, rather than to a different type that is expected to benefit greatly, that suggests the policy implicitly places higher welfare weight on the first type. Or, if a policy consistently allocates to applicants whose health improves as a result of the intervention --- instead of applicants whose consumption increases --- that implies the policy implicitly highly values health.

To illustrate how this method can be used to interrogate a real-world policy, we apply it to historical data from PROGRESA, one of the world's largest (and best-studied) anti-poverty programs.
% which provided cash transfers to eligible households in Mexico.\footnote{PROGRESA conditioned payouts on certain actions, but we treat the program as an unconditional transfer. For simplicity, we assume PROGRESA did not have differential spillover benefits on different households.} 
In this context, we first estimate the heterogeneous treatment effects of the program using \citeauthor{wager_estimation_2018}'s \citeyearpar{wager_estimation_2018} causal forest machine learning method (though as we discuss, alternative methods for estimating treatment effects would work as well). Consistent with prior work, we find evidence of treatment effect heterogeneity --- for instance, that indigenous households benefit most from the program \citep[cf.][]{djebbari_heterogeneous_2008}. 

We then use our method to estimate the preferences consistent with the observed ranking of households and its heterogeneous effects on consumption, child health, and school attendance. %We focus on the three outcomes prioritized by the program (consumption, child health, and school attendance).
We find that indigenous households were more likely to be allocated the program, but because they benefit so much more, that the policy is actually consistent with weighting them 11.7\% \emph{lower} than non-indigenous households. 
Our results also suggest that the program's design is consistent with assigning extra value to poorer, larger, and less educated households: households are weighted 0.14\% lower for each 1\% increase in household income, 5.7\% higher for each additional person in the household, and 32.8\% lower if the household head has a high school education. These valuations, estimated using our method, are similar to the stated preferences of Mexican residents, as measured by hypothetical allocation questions in a survey we conducted in 2021. We additionally recover estimates of how the policy implicitly values impacts on health and schooling relative to consumption. Confidence intervals are imprecise but admit conventional valuations.

%, households with more members, and non-indigenous households. 
% For instance, the estimates imply that households are weighted 11.7\% \emph{lower} if indigenous, 0.14\% lower for each 1\% increase in household income, 5.7\% higher for each person in the household, 1\% lower for each of the household head's year of age, and 32.8\% lower if the household head has a high school education. %\footnote{We compute the average percentile change by first computing how each
% household's projected ranking would shift, given different covariates,
% and then taking the median change over all households.} 
%Our estimates of weights on different impacts are imprecise, but 95\% confidence intervals suggest that the program valued each additional day of school attendance at less than a 36.3\% of a day of consumption, and valued each averted sick day at less than 1.8\% of a day of consumption. 

%To benchmark these results, we compare the implied preferences estimated using our method to the stated preferences of Mexican residents, as measured by hypothetical allocation questions in a survey. These preferences are imprecisely measured, but we find similar welfare weights on income, and confidence intervals overlap for the valuation of the other outcomes.

Finally, we show how this approach can further be used to evaluate counterfactual policies and preferences. In the PROGRESA case, we show what  \textit{would have occurred} had the program designers placed higher value on certain types of impacts (e.g., health vs. education) or certain types of households (e.g., equal welfare weights). This analysis suggests that, for instance, a policymaker who cared exclusively about impacts on schooling should prefer a policy that prioritizes richer households; a policymaker that valued only health impacts would instead prioritize indigenous households. More broadly, we show where these counterfactual policies lie relative to the Pareto frontier that characterizes improvements across the three focal welfare outcomes.
% If the
% government assigned welfare weight solely based on preferring lower income households, the rule would
% slightly de-prioritize indigenous households and households with small
% children. 
% A technocratic ranking that weighs impacts
% according to external cost benefit estimates would have traded
% off increased sick days to reduce missed school days and increase
% consumption. 
%In practice, implemented policies may balance the needs of multiple advocates. We can use our method to assess those contributions. An education minister who valued only schooling impacts would advocate for increasing the priority on households of indigenous status. A minister valuing only health impacts would advocate for increasing the priority of smaller households. A minister valuing only consumption impacts would advocate to slightly increase the priority of indigenous households and households with lower income. Finally, we assess the impact of all of these alternative allocations on consumption, sick days, and missed days of school. We show where these allocations lie relative to the Pareto frontier of possible outcomes.

Taken as a whole, this approach makes it possible to invert the discussion about policies and programs. Rather than debate the means of the policy (who is eligible, how large are the benefits?), this framework makes it possible to debate the ends (how much do we value health, education, or consumption? Should poor families be prioritized over middle class families?). Indeed, modern econometric methods have begun to reveal that many policies benefit some groups more than others \citep[cf.][]{haushofer_targeting_2022}; our framework suggests how policies might be reconciled with that heterogeneity.
The framework can be applied to a wide range of settings where policymakers allocate scarce resources and heterogeneous treatment effects can be estimated. %Our method does not require access to all of the information that policymakers use to develop a ranking; it can be computed with the final ranking and welfare relevant characteristics of individuals.

The approach has three caveats. First, it requires defining which outcomes and characteristics are allowed to enter the objective function; outcomes that are not included are assumed to not be valued. While this decision is a substantive one, the method is sufficiently flexible to allow for other definitions of welfare. Second, in order to estimate how different types of people are affected by the intervention, the analyst must observe experimental variation in access to the intervention for all types of people (including both those who are ultimately eligible and ineligible under the policy). This is commonly the case with randomized controlled trials or pilots. Third, it requires
a large enough dataset to estimate both heterogeneous treatment effects and the implied welfare parameters. These datasets are increasingly becoming available, particularly in settings with digital experimentation.

\subsection*{Related Literature}

This paper contributes to literature on optimal targeting and taxation
\citep{nichols_targeting_1982,barr_economics_2012,fleurbaey_optimal_2018},
including work comparing targeted policies to universal basic income \citep{alatas_targeting_2012,hanna_universal_2018}.
It can be viewed as a response to \citet{ravallion_how_2009}, which argues that targeting poverty directly may not be sufficient for impact, and suggests that it may be better to target based on desired outcomes. In that sense, our work relates closely to \citet{haushofer_targeting_2022}, which asks how targeting on treatment effects compares to targeting on baseline poverty. Their empirical analysis suggests that those who are most impacted by a Kenyan cash transfer are not always the poorest. Our paper focuses on the inverse problem of estimating the welfare function consistent with an observed policy. The two approaches are thus complementary; our also extends from a specified utility function defined over a single outcome to a general welfare function that can rationalize targeting based on household characteristics as well as impacts on multiple outcomes. 
%(in that, it relates to work that infers policymaker preferences from their actions \citep{timmins_measuring_2003}). 
Our empirical results also engage with research on the effects and allocation of cash transfer programs \citep{behrman_randomness_1999,skoufias_targeting_2001,gertler_conditional_2004,john_hoddinott_impact_2004,coady_welfare_2006,djebbari_heterogeneous_2008,alderman_contribution_2019}. We build on this work by showing how effects can be used to audit policymaker priorities, and improve the design of
future policies. 

Our approach also relates to a growing literature that takes a given welfare function as fixed, and considers what are the best decisions to take. \citet{kitagawa_who_2018} computes optimal assignment of treatment with experimental data, and \citet{athey_policy_2020} with observational data. \citet{gechter_evaluating_2019} assesses how well different ex ante treatment assignments maximize a given welfare function under ex post experimental data. \citet{wang_optimal_2020} considers the theoretical problem of allocating resources given heterogeneous aid agency preferences over individuals, and describes allocation queues as a solution to a combinatorial problem. This literature faces a central problem: what notion of welfare do, or should, societies maximize? Our paper takes a step towards answering this question, by solving the reverse problem: estimating welfare functions consistent with observed decisions.

% Our approach relates to the index problem of measuring welfare \citep{rawls_theory_1971}. 
It is increasingly common to construct indices summarizing multiple outcomes as a more nuanced measure of welfare \citep{greco_methodological_2019}. A persistent question in assembling these indices is what weight to apply to each component. These weights have economic meaning: how valuable is one component relative to another? Common approaches are geometric: setting equal values to each component \citep{undp_human_1990}, or analyzing how components vary together in observational data, using a principal component analysis \citep{filmer_estimating_2001,mckenzie_measuring_2005}. We derive weights that have an economic interpretation using revealed preferences, how policies implicitly make trade-offs. A related approach is to set weights to optimally predict some gold standard measure of utility, if one is available \citep{jayachandran_using_2021}. 
%These approaches have a geometric, rather than economic interpretation; while they are helpful for describing observed patterns, it is unclear whether they should be used to measure impacts: the welfare benefit of 1 unit increase in a geometric index will differ, depending on which components it results from. 
% Many measures of welfare derive an index out of subcomponent measures (for example, Disability Adjusted Life Years, Womens' Empowerment, etc.). 
% One could set the weights equally (as in PHQ-9). A common way to set these weights is by analyzing how components vary together in observational data, using a principal component analysis \citep{ewerling_swper_2017}. This can lower the sampling noise in the combined measure, but these weights have a statistical, not economic interpretation. There is no strong reason that societies should place more value on the components that are weighted more highly by a PCA; as a result, the welfare benefit of 1 unit increase in a PCA index will differ, depending on which components it results from. 

Also related is a recently expanding public finance literature on
welfare weights. \citet{hendren_efficient_2019} infers the weight
on different households implied by a tax schedule, based on the distortions
required to transfer them resources. \citet{saez_generalized_2016}
generalize welfare weights to reconcile popular notions of fairness
with optimal tax theory. Our paper shows how similar welfare questions
can be raised across a broad set of domains where heterogeneous treatment
effects can be estimated.

More broadly, our efforts also connect with recent computer science scholarship on fairness in machine learning \citep[cf.][]{dwork_fairness_2012,barocas_fairness_2018}.
Several papers in this literature study the social welfare implications of algorithmic decisions, and how social welfare concerns relate to different notions of fairness \citep{ensign_runaway_2017,hu_welfare_2018,mouzannar_fair_2018,liu_delayed_2018}. This relates to work on multi-objective machine learning \citep{rolf_balancing_2020}. %develop methods for training ML algorithms to optimize multiple objectives); 
\citet{kasy_fairness_2020} describe limitations of fairness constraints, and suggest that algorithms should be optimized for impacts. %TThat paper suggests theoretically that the welfare weights implied by an algorithm's allocation could be used as a measure of power - but this stops short of allowing for prioritization across different measures of welfare, and is not demonstrated empirically. 
Also related, \citet{noriega_algorithmic_2018} discuss how
different constraints to targeting can impact efficiency and fairness.
Our approach is distinct, however, in that we show how using machine
learning tools can be used to better characterize and audit the values consistent with a program's observed allocation.
We hope that by providing increased visibility into these revealed
preferences, future policies can be better aligned with stated preferences
and explicit policy objectives.

\section{Model}

\setcounter{subsection}{0}

We consider the problem of allocating treatment among $N$ entities, which could be, for example, individuals, households, firms, or regions. For convenience, we refer to entities as households.

A policymaker has ranked each household $i$ in the priority order they will be allocated some benefit or treatment, $T_{i}\in\{0,1\}$. This ranking $z_{i}$ may include ties between households; in the extreme it could simply represent the binary decision of whether household $i$ will be allocated treatment ($z_i\in \{0,1\}$).

We attempt to reconcile that ranking with an implicit welfare function:
\[
S=\sum_{i}S_{i}
\]
\[
S_{i}=\mu(\mathbf{x}_{i})\cdot u_i(T_{i})
\]
where each household $i$ is valued according to some objective utility $u_i(T_i)$, scaled by some differential welfare weight $\mu(\mathbf{x}_{i})$ based on its characteristics $\mathbf{x}_{i}$, with a functional form to be specified later. Objective utility $u_i$ can be decomposed into components:
\[
u_i(T_{i})=v_{i0}(T_{i})+\sum_{j>0}\lambda_{j}(\mathbf{x}_{i}) v_{ij}(T_{i})+C\cdot T_{i}
\]
where $v_{ij}$ represents a component of utility (such as consumption, or health), and $\lambda_{j}(\mathbf{x}_{i})$
represents $j$'s implied value relative to the numeraire or reference outcome ($j=0$), with a functional form to be specified later. $C$ is a constant representing
the net intrinsic value of providing the program, even absent impact.\footnote{For intuition: if $C$ is very large, the ranking between households is explained mostly by differences in welfare weights; if $C$ is small or zero, the ranking depends also on impacts.}

Imagine we knew the impact of treatment on household $i$'s component of utility $j$: $\Delta v_{ij}\coloneqq v_{ij}(1)-v_{ij}(0)$. The welfare impact of treating household $i$ could then be written
\[
\Delta S_{i}=\mu(\mathbf{x}_{i})\cdot\left(\Delta v_{i0}+\sum_{j>0}\lambda_{j}(\mathbf{x}_{i})\Delta v_{ij} + C\right)
\]

The ranking could then be reconciled with ordering households according to their implied welfare impact from receiving treatment,
\begin{equation}
z_{i}=f(\Delta S_{i}+\epsilon_{i})\label{eq:ranking}
\end{equation}
where $f$ is a weakly increasing transformation, which preserves the ranking of households. The shock $\epsilon_{i}$ may represent measurement error in estimates of welfare, or mistakes in the allocation.

\subsection{Measuring Utility Impacts\label{subsec:MeasuringUtility}}

Reconciling the observed ordering of households with the welfare impacts of the policy requires that we have an estimate of the impact of treatment on utility for each household. We will assume that each utility component $v_{ij}$ is a function of an observed outcome $y_{ij}$,
\[
\Delta v_{ij} \coloneqq v_{ij}(T_i=1)-v_{ij}(T_i=0) = g_{j}(y_{ij}^1) - g_{j}(y_{ij}^0 )
\]
where $g_j$ represents the utility function for $j$ (which could be, for example, $g_j(y)=\log(y)$, or $g_j(y)=y$).%
\footnote{We assume that these functional forms are known, but they could be estimated within our setup. If the $g_j(\cdot)$ utility functions are incorrectly specified to be linear, then welfare weights $\mu(\mathbf{x}_i)$ and the vector of impact weights $\boldsymbol{\lambda}(\mathbf{x}_i)$ can measure the combination of the underlying welfare weights and curvature in utility to first approximation. See Appendix Section \ref{subsec:generalizedcurvature}.}  Additionally, we assume that we have an experimental design that makes it possible to predict the heterogeneous effects of treatment on each household and each outcome $\Delta \hat{v}_{ij}$.

\subsection{Intuition}

\begin{figure}
\begin{tabular}{ll}
\multicolumn{2}{c}{\textbf{An allocation rule that prioritizes the poor} (low $x_i$)}\\
\multicolumn{2}{c}{\includegraphics[width=5cm]{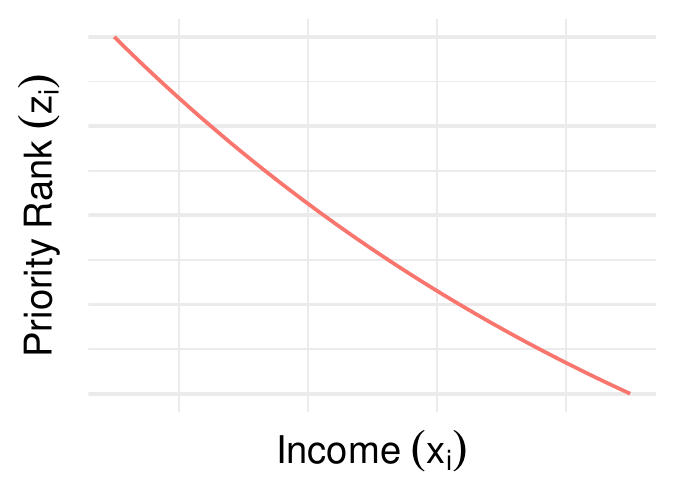}}\\
\multicolumn{1}{l}{\textbf{Could result from}} & \\
(1) Higher welfare weight on the poor & if treatment effects are constant\\
\includegraphics[width=5cm]{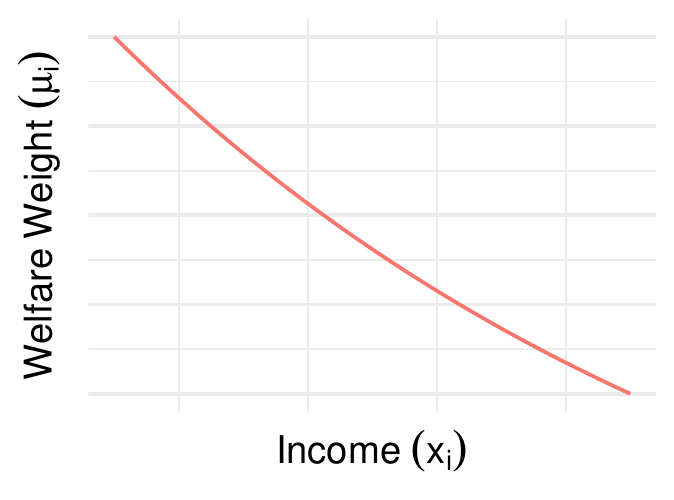} & \includegraphics[width=5cm]{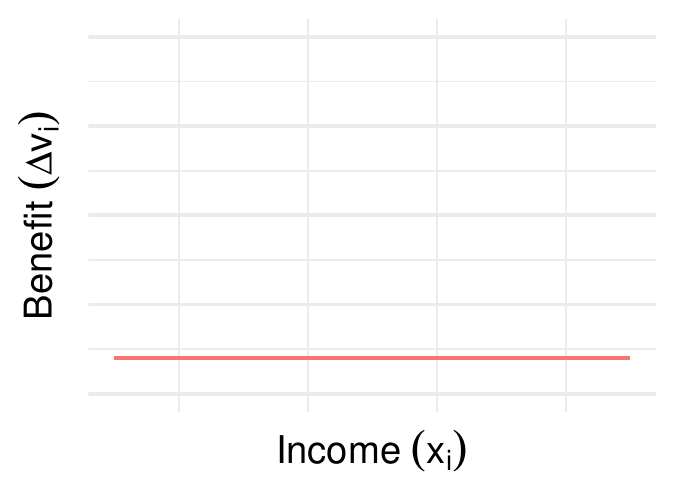}\\
(2) Equal welfare weights on households & if treatment effects are higher for the poor\\
\includegraphics[width=5cm]{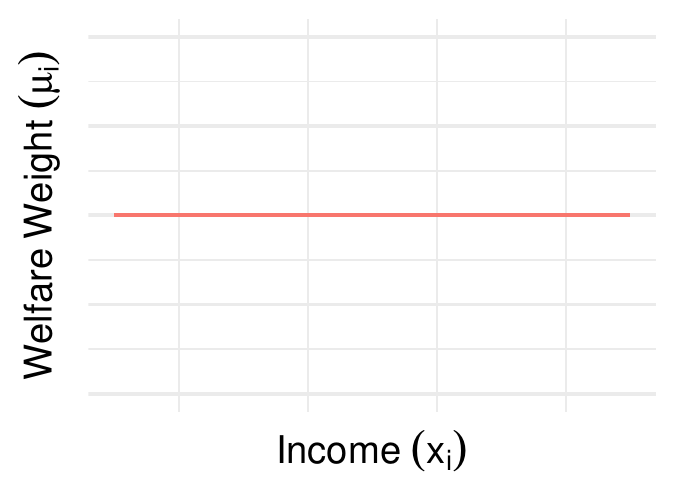} & \includegraphics[width=5cm]{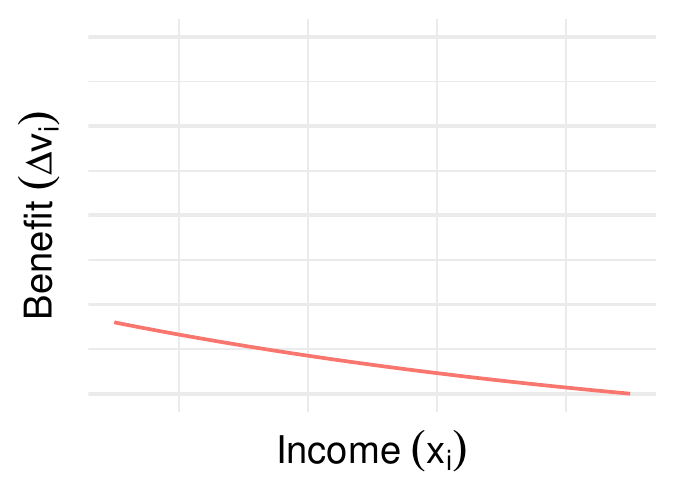}\\
(3) Higher welfare weight on the rich & if treatment effects are much higher for the poor\\
\includegraphics[width=5cm]{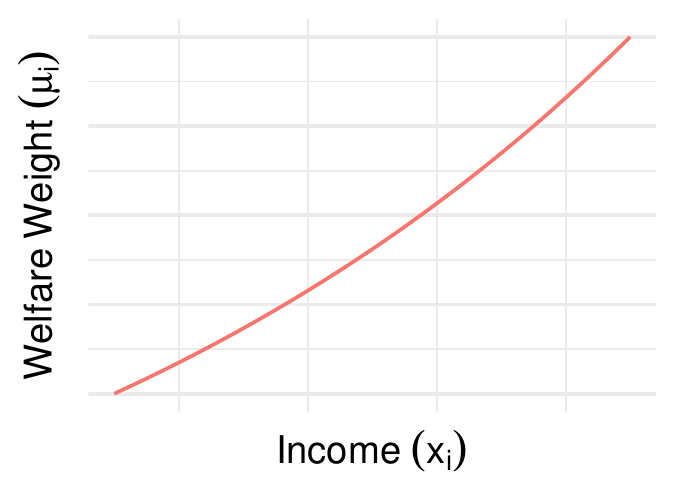} & \includegraphics[width=5cm]{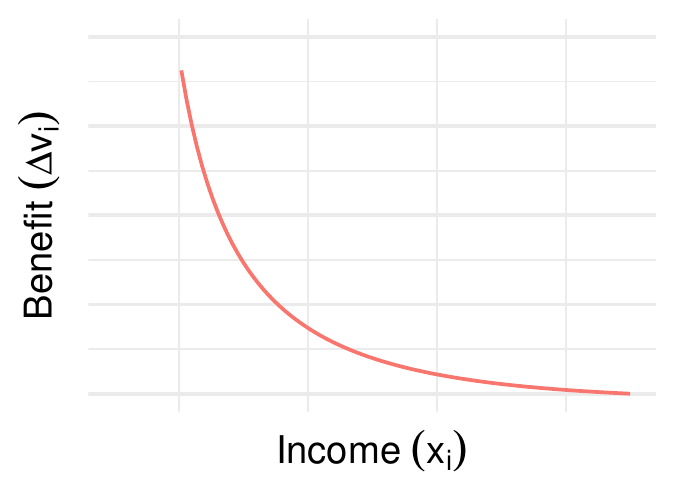}\\
\end{tabular}
\caption{Intuitive Example\label{fig:Intuitive-Example}}
\end{figure}

To demonstrate the intuition behind our method, we illustrate with a simple
example in Figure \ref{fig:Intuitive-Example}. Consider the case of a single outcome and one dimension of heterogeneity, $x$, which
corresponds to income. A policymaker allocates a program by
ordering households by the function $z(x)$, prioritizing poor households.
As shown in Figure \ref{fig:Intuitive-Example}, depending on how treatment effects vary with $x$, the same allocation could result from (1) higher welfare weights on the poor, (2) equal welfare weights, or (3) higher welfare weights on the rich.
Likewise, in the case where $x$ is binary, an allocation to one group can result from (i) higher welfare weights, if that group benefits the same or less; (ii) equal welfare weights, if that group benefits more; or (iii) lower welfare weights, if that group benefits much more.

%Alternately, in the case where $x$ is binary,
%\[
%\resizebox{1.0\hsize}{!}{
%\textnormal{An allocation to one group can result from }\begin{cases}
%\textnormal{higher welfare weights} & \textnormal{if that group %benefits the same or less}\\
%\textnormal{equal welfare weights} & \textnormal{if that group %benefits more}\\
%\textnormal{lower welfare weights} & \textnormal{if that group benefits much more}
%\end{cases}
%}\]

The next section demonstrates how to empirically recover welfare and
impact weights from data in when there are multiple dimensions of
heterogeneity and multiple outcomes of interest.

% \begin{figure}
% % \includegraphics[width=7in]{./Code/figs/simulations/simulation}
% \centering
% \includegraphics[width=5.5in]{./Code/figs/simulations/simulation}
% \caption{Intuitive Example\label{fig:Intuitive-Example}}
% \end{figure}
% \vfill{}

\vfill{}
\clearpage\pagebreak{}

\section{Estimation}
\label{sec:estimation}
\setcounter{subsection}{0}

Estimation proceeds in two steps:

We first predict the effect of treatment on each outcome $\Delta\hat{v_{ij}}=\Delta \hat{v}_{j}(\tilde{\mathbf{x}}_{i})$, which may
be heterogeneous as a function of covariates $\tilde{\mathbf{x}}_{i}$. This can be done using a variety of methods for estimating heterogeneous treatment effects, such as \citeauthor{wager_estimation_2018}'s (\citeyear{wager_estimation_2018}) machine learning approach, which works well when there is experimental variation in treatment assignment.%
\footnote{For treatment effect heterogeneity, covariates need only be correlated with impacts, so one may wish to include a large set of covariates in $\tilde{\mathbf{x}}_i$. In contrast, the choice of $\mathbf{x}_i$ determines which characteristics to allow favoritism over, so one may wish to justify a smaller set.} 
% We can use this to predict the outcome for both the factual and counterfactual state.

% If $i$ is control ($T_{i}=0$):
% \begin{align*}
% \hat{v}_{ij}^{0} & =v_{ij}\\
% \hat{v}_{ij}^{1} & =v_{ij}+\Delta \hat{v}_{j}(\tilde{\mathbf{x}}_{i})
% \end{align*}

% or treated ($T_{i}=1$):
% \begin{align*}
% \hat{v}_{ij}^{0} & =v_{ij}-\Delta \hat{v}_{j}(\tilde{\mathbf{x}}_{i})\\
% \hat{v}_{ij}^{1} & =v_{ij}
% \end{align*}

% Here we have assumed that treatment effects are known with certainty; in the estimation section we will consider adjustments for uncertainty.

Then, given that the policy achieves effects estimated to be $\Delta\hat{v_{ij}}$, we ask what preferences (i.e., $\mu(\mathbf{x}_{i}) $, the vector $\boldsymbol{\lambda}(\mathbf{x}_{i}) $, and $C$)
would be consistent an allocation according to the given ranking ($\boldsymbol{z}$)?
If household $i$ is prioritized over $i'$ ($z_{i}>z_{i'}$),
equation~\eqref{eq:ranking} implies:
\begin{eqnarray*}
 & \mu(\mathbf{x}_{i})\cdot\left(\Delta\hat{v_{ij}} +\sum_{j>0}\lambda_{j}(\mathbf{x}_{i})\Delta\hat{v_{ij}}+C\right)+\epsilon_{i}> %\\
%  & >\\
%  & 
 \mu(\mathbf{x}_{i'})\cdot\left(\Delta\hat{v_{i0}}+\sum_{j>0}\lambda_{j}(\mathbf{x}_{i'})\Delta\hat{v_{ij}}+C\right)+\epsilon_{i'}
\end{eqnarray*}

This problem can be modeled with an ordinal logit likelihood if we make the common assumption that the ranking error is distributed extreme value type-I: $\epsilon_{i}\sim\sigma\cdot EV(1)$. Then, the logit likelihood of this particular placement of $i$ in the ranking $\boldsymbol{z}$ is:

\begin{equation}
    l_{i}=\frac{\exp\left[\frac{1}{\sigma}\cdot\mu(\mathbf{x}_{i})\left(\Delta\hat{v_{i0}}+\sum_{j>0}\lambda_{j}(\mathbf{x}_{i})\Delta\hat{v_{ij}}+C\right)\right]}{\sum_{i'\epsilon\Lambda_{i}}\exp\left[\frac{1}{\sigma}\cdot \mu(\mathbf{x}_{i'})\left(\hat{\Delta v_{i'0}}+\sum_{j>0}\lambda_{j}(\mathbf{x}_{i'})\hat{\Delta v_{i'j}}+C\right)\right]}
    \label{eq:likelihood_i}
\end{equation}
where $\Lambda_{i}=\{i'|z_{i'}<z_{i}\}$ represents the set of households ranked lower than household $i$. 

The logit likelihood of the full observed ranking $\boldsymbol{z}$
is therefore
\[
L(\boldsymbol{z},\mathbf{x},\boldsymbol{\mu},\boldsymbol{\lambda},C,\sigma)=\prod_{i}l_{i}
\]

We use maximum likelihood to estimate the functions $\mu(\mathbf{x}_{i})$ and $\boldsymbol\lambda(\mathbf{x}_{i})$ and parameters $C$ and $\sigma$ that best match the observed data $\{\boldsymbol{z},\mathbf{x},\{\Delta\hat{v_{ij}}\}_{ij}\}$. Unlike standard discrete choice settings where partial orderings are observed for multiple decisionmakers, we observe a single ordering of all alternatives. For this type of ranked data, we follow the exploded logit likelihood described by \cite{train_discrete_2009}.%
% \footnote{The likelihood considers the selection of the highest ranked household against all alternative households. Then, the highest ranked household is dropped from the comparison, and the likelihood considers selection of the second highest ranked household against all lower ranked households. It continues in this manner through to the lowest ranked household.}

% For components of utility with $g_j(\cdot)$ that is linear in its underlying outcome, we can simply compute $\Delta\hat{v_{ij}}=g_{j}(\hat{y}_{ij}^1)
% - g_{j}(\hat{y}_{ij}^0 )$ as shown in Section \ref{subsec:MeasuringUtility}. However, this expression would be a biased estimate of the change in that utility component if there is curvature in $g_j(\cdot)$ and uncertainty in $\hat{y}_{ij}^{T}$, due to Jensen's inequality. For utility components $j$ with curvature, we compute $\Delta\hat{v_{ij}}=\frac{1}{D}\sum_{d=1}^{D} \left[ g_{j}(\hat{y}_{ijd}^1) - g_{j}(\hat{y}_{ijd}^0 ) \right]$ for $D$ bootstrapped estimates of treatment effects to obtain an unbiased estimate.

Standard errors are computed by bootstrapping the entire procedure (accounting for uncertainty in both treatment effects and preference parameters). Individuals are drawn with replacement, and these bootstrapped samples are used to compute treatment effects, and then welfare and impact weights.
% Standard errors reported are the standard deviation across bootstrapped welfare and impact weight estimates.

In many settings, we may not observe a full ranking or score, but rather a binary allocation of beneficiaries and non-beneficiaries ($T_{i}\in\{0,1\}$). This corresponds to a ranking with two levels. In such settings, the same procedure described above can be applied.

\subsection{Parameterization}

Our framework will work with general functional forms for $\mu(\mathbf{x}_{i})$ and $\lambda_j(\mathbf{x}_{i})$. In the empirical application that follows in Section~\ref{sec:application}, we model welfare weights as multiplicative:
\[
\mu(\mathbf{x}_{i})=\ensuremath{{\Pi}}_k\omega_k^{x_{ik}}
\]
We model the relative weight on outcome $j$ as the same for all households, $\lambda_{j}(\mathbf{x}_{i}) \equiv \lambda_{j}$, because our sample is not large enough to differentiate heterogeneity on these dimensions.

\subsection{Identification}

Preferences are identified based on how the policy's ranking ($z_{i}$) varies with characteristics ($\mathbf{x}_{i}$) and with treatment effects on components of utility ($\Delta\hat{v_{ij}}$).

\paragraph{Unobservables}
Our approach estimates the preferences that are consistent with the implemented policy $z_{i}$, given the estimates of impact $\Delta\hat{v_{ij}}$. This can be thought of as an ex-post audit. Our estimates will recover an observed component of welfare, $\Delta S_{i}$, that is uncorrelated with any unobserved component, $\epsilon_{i}$. There are several reasons why these implied preferences might differ from the actual preferences of the policymaker.

First, implied preferences could differ from policymaker preferences if the policymaker based the ranking on correlated unobservables. For example, if a policymaker is racially biased but an analyst does not allow race to enter modelled preferences, the policy may be found to be consistent with a preference for an income level that is correlated with race. In such settings, the method still reveals preferences that are \textit{consistent} with the policy's values, under the given specification of preferences.%
\footnote{This is analogous to the way that ordinary least squares recovers the best linear predictor given included variables, even in the presence of omitted variables.} 
The specification of preferences (i.e., which variables they are defined over and their functional form) is thus a substantive decision. For this reason, practical applications should include both characteristics that policymakers wish to prioritize as well as characteristics for which there may be concerns of bias. % In our application, we select relevant characteristics based on a survey of constituents. Treatment effects should be measured on the outcomes that are most relevant for welfare, as impact evaluations are typically designed to do.

Second, implied preferences could differ from policymaker preferences if the policymaker has incorrect beliefs about these impacts at the time of the decision. If that were the case, upon observing the results of our method, the policymaker could change the policy to better align with their preferences. The method thus provides a tool for course correction.

The method can also be applied in cases where there is no single policymaker\textemdash for example, where allocations are the result of deliberations between constituents. In that case, our method will reveal social preferences consistent with the final allocation.

\paragraph{Sufficient variation}
Identification also requires sufficient variation. It requires that some households benefit more than others. Welfare weights $\boldsymbol{\omega}$ are primarily identified based on heterogeneity in impacts on the numeraire utility $\Delta\hat{v_{i0}}$. If treatment effects were homogenous, it would not possible to separately identify $\boldsymbol{\omega}$ and $\boldsymbol{\lambda}$ (their combination may be identified, in which case our method would collapse down to a standard exploded logit that does not account for treatment effects). 

Identification of $\boldsymbol{\lambda}$ also requires that treatment has different impacts on different components of utility. Impact weight $\lambda_j$ is identified from the relative ranking of households that are impacted more or less on utility component $j>0$ than on the numeraire ($j=0$). If the treatment effects were heterogeneous but colinear between different components of utility, it would be possible to identify $\boldsymbol{\omega}$ but not $\boldsymbol{\lambda}$, because the data would not reveal how different components of utility influence the ranking.

The resulting parameters $\boldsymbol{\omega}$ reveal which characteristics $\mathbf{x}$ are correlated with being prioritized. If $\mathbf{x}$ includes both a relevant variable $x_{ik}$ as well as an irrelevant but colinear variable $x_{ik'}$, the method will have imprecise estimates of the contribution of both, in a similar manner as a standard regression would. In that sense, one may want to restrict analysis to characteristics $\mathbf{x}$ that one believes may be relevant for differential preference. In our application, we use survey data to narrow down factors that should enter the targeting rule.

%In all of these cases, lack of identification can be recognized from having large standard errors.

\section{Application}
\label{sec:application}

To illustrate how our method can be used in applied settings, we use the case of PROGRESA, a large conditional cash transfer program in Mexico.

\subsection{Background on PROGRESA}
First implemented by the Mexican federal government in 1997, PROGRESA provided cash transfers to poor households. Transfers, which averaged 197 pesos per month (approximately \$20 USD at the time), were conditioned on regular doctor's visits and/or regular school attendance \citep{john_hoddinott_impact_2004}. Roughly 99\% of enrolled households met these conditions \citep{simone_boyce_experiment_2003}.\footnote{For simplicity, our analysis does not account for the conditionality of the transfer. For a more detailed discussion of PROGRESA and its background, see \citet{emmanuel_skoufias_conditional_2008}, and \citet{simone_boyce_experiment_2003}.}

%\subsubsection*{Targeting}

% multidimensional poverty, disciminant. income per capita, within each region, find variables that discriminate between poor and nonpoor
% The discriminant analsysis allowed regional differences but \citet{skoufias_evaluation_1999} 
Within poor communities, PROGRESA ranked households based on a `household poverty score' that incorporated a variety of different characteristics (such as household structure, indigenous languages, occupation, income, housing materials, etc.).%
\footnote{The program defined poor communities as those with a high `village marginality index', computed based on the proportion of households living in poverty, population density, and health and education infrastructure. We focus on the preferences implied by household poverty scores, which were the basis for determining which households within a community eligible for the program.} 
The score was computed in three steps. First, each household was classified as poor or not poor based on per capita income. Second, that poverty classification was approximated using discriminant analysis based on household characteristics \citep{skoufias_evaluation_1999}. Third, the list of eligible households was presented in meetings in each community for review; a small number of households changed classification as a result. Our focus is on understanding which underlying values are consistent with the allocation resulting from this method of determining eligibility for the program.

%\citet{skoufias_evaluation_1999} finds that the resulting ranking is substantially different from consumption-based targeting; however we could find little documentation that describes the rationale for these deviations. 

%\subsubsection{Experimental Design}

During its initial implementation, PROGRESA administrators used a staggered roll-out to randomize when villages could enroll in the program: of the 506 villages included in the evaluation, 320 were randomly assigned to treatment, and initiated
into the program in summer 1998. 186 communities were assigned to
control and were not initiated into the program until 2000. \citet{behrman_randomness_1999}
show that the randomization across communities was successful in that
treatment and control communities were statistically indistinguishable
across a wide array of observable covariates.

\subsubsection*{Data}

Our analysis relies on two distinct sources of data. The first is a standard household survey conducted in October 1998 (baseline) and November 1999 (endline). These capture household demographics, socioeconomic characteristics, health care utilization, and educational attendance for 14,333 households over the entire experiment period -- see Appendix Table \ref{tab:Descriptive-Statistics} for summary statistics. We focus on the sample of 6,642 households over which our outcomes are defined, who have at least one child aged 5 or below and at least one child aged 6-16. Thus, our estimates will reveal the values implied by the rankings within households with children, and not between households with and without children in the relevant ages.

The second data source is a survey that we conducted in 2021 to understand the preferences of Mexican residents over how households should be prioritized for social assistance. We surveyed a sample of 315 Mexican residents to elicit preferences for which types of households should receive transfers, and what types of program impacts were most desirable, in a manner similar to \citet{saez_generalized_2016}. The survey asked respondents which household attributes should be considered in the design of such a program, and relied on multiple price lists to elicit indifference points. For a complete description of this survey, see Appendix~\ref{subsec:preference-survey}.

%\subsubsection{Variables}
We focus on three welfare outcomes that were monitored in the household surveys: (i) logarithm of \textit{per-capita consumption}; (ii) \textit{child health}, measured as the average number of sick days
per child aged 0-5; and (iii) \textit{school attendance}, calculated as the average number of school days missed per child aged 6-16. We treat log consumption as our numeraire ($g_0(y_{consumption})=\log(y_{consumption})$), and allow the other two outcomes to enter the welfare function linearly ($g_j(y_j)=y_j$ for $j>0$).%
\footnote{\citet{gandelman_risk_2015} fails to reject a level of risk aversion consistent with logarithmic utility in Mexico, based on self-reported wellbeing.} 
Previous studies have estimated significant treatment effects on all three outcomes using the same survey data \citep{john_hoddinott_impact_2004,emmanuel_skoufias_conditional_2008,simone_boyce_experiment_2003,djebbari_heterogeneous_2008}. Note that the program could also have impacted other outcomes not measured; our method will assume that such impacts are either zero or not valued. In Section~\ref{sec:implied_prefs}, we discuss implications and extensions of this simplifying assumption.

We define welfare weights $\mu(\mathbf{x}_i)$ over the top five characteristics that Mexican residents in our survey reported should be considered when targeting cash transfers ($\mathbf{x}_i$): income; number of people; and age, education, and indigenous status of the household head.

\subsection{Characterizing the Decision Rule}

As a first step, we characterize the decision rule, by indicating which types of households are observed to be ranked higher than others. Table~\ref{tab:Allocation-Rule-Actual} column 1 reports these results, where the contribution of household characteristics to the final ranking $\boldsymbol{z}$ is estimated with a logit ranking model (i.e., our model's likelihood equation~\eqref{eq:likelihood_i} with constraints $\Delta\hat{v}_{ij}\equiv 0$ and $C=1$, estimating the constrained weights $\tilde{\boldsymbol{\omega}}$). We report coefficients transformed by log base 1.01 ($\log_{1.01}(\tilde{\boldsymbol{\omega}})$), which can be interpreted as the number of successive 1\% increments implied. This suggests that households that are indigenous are ranked 47\% higher ($1.01^{38.6}$). It also suggests that each 10\% increase in income corresponds with a 2\% decrease in ranking. Each additional household member is associated with a 14\% ($1.01^{13.0}$) increase in ranking. However, the conventional regression in column 1 does not describe \textit{why} these households are ranked highly; it could be that they benefit more (higher treatment effects) or that they are favored (higher welfare weights), as suggested in Figure~\ref{fig:Intuitive-Example}.

%%%%%%%%%%%%%%%%% TABLE  %%%%%%%%%%%%%%%%%%%%%%%%%%%%%%
\begin{table}
\centering \caption{What Values are Consistent with the PROGRESA Decision Rule?\label{tab:Allocation-Rule-Actual}}
\renewcommand{\arraystretch}{1.2}
\begin{threeparttable} 
%\begin{footnotesize}
\begin{tabular}{lccc}
\toprule
 & \multicolumn{2}{c}{Household Poverty Score 1999}\\
 & \textbf{Decision Rule}   & \textbf{Implied Preferences} \\[.2em]
& (Prioritization)  & Welfare Weights $log_{1.01}(\boldsymbol{\omega})$ \\
\cline{2-3}
\emph{(number of 1\% increments)} &  &   \\
{Indigenous}  & { 38.57 (5.0)}  &  { -12.4 (4.2)}  \\
{log(Income)}  & { -24.9 (1.9)}   & { -14.3 (5.4)}  \\
{Household Size}  & { 13.0 (0.8)}   & { 5.6 (2.1)}  \\
{Head Age}  & { -2.31 (0.2)}  &  { -1.0 (0.6)}  \\
{Education (HS or above)}  & { -240.2 (848.5)}   & { -39.9 (21.5)}  \\ [.6em]
 &  & Impact Weights \\
\cline{2-3}
\emph{(log points of daily consumption)} &  &   \\
{Missed Schooling (per day)}  &  \textbf{ $\lambda_{1}$}  & { -0.03 (0.17)}  \\
{Sickness (per child sick day)}  &  \textbf{ $\lambda_{2}$}  & { 0.08 (0.05)}  \\
{Value Regardless of Impact}  & { $C$}  & { 0.47 (3.75)}  \\[.4em]
{$\sigma$}  &  &  { 0.17 (0.17)}\\[.4em]
$N$  & 6642 & 6642 \\
\bottomrule
\end{tabular}
%\end{footnotesize}
\begin{tablenotes}[normal,flushleft] 
\footnotesize
\item \emph{Notes:}
Left column is computed using our method, without treatment effects included in the estimation. Right column is calculated using causal forests to estimate heterogeneous treatment effects (see Figure~\ref{fig:Distribution-of-Estimated-CF}). In all columns, standard errors are computed using a two-step bootstrap procedure that accounts for uncertainty in both treatment effects and preference parameters. Observations are drawn with replacement before estimation of the treatment effects and the welfare and impact weights. Treatment effects are then estimated from these bootstrapped samples, and welfare and impact weights estimated from these bootstrapped treatment effect estimates; the standard errors reported are the standard deviation across bootstrapped welfare and impact weight estimates.
\end{tablenotes} 
\end{threeparttable} 
\end{table}
%%%%%%%%%%%%%%%%%%%%%%%%%%%%%%%%%%%%%%%%%%%%%%%

\subsection{Results: Estimating What Policies Value}

Our main results from the PROGRESA example show how our method can decompose an observed allocation into the values implied by the decision rule.

\subsubsection{Heterogeneity in Treatment Effects}

As has been documented in prior work, the PROGRESA program significantly increased household welfare. On average over our sample, PROGRESA increased log household monthly consumption by 0.135, reduced the number of sick days per child by 0.18, and reduced the number of school days missed per child by 0.005.

\begin{figure}
\caption{Distribution of Estimated Treatment Effects\label{fig:Distribution-of-Estimated-CF}}
\begin{center}
\includegraphics[width=\textwidth]{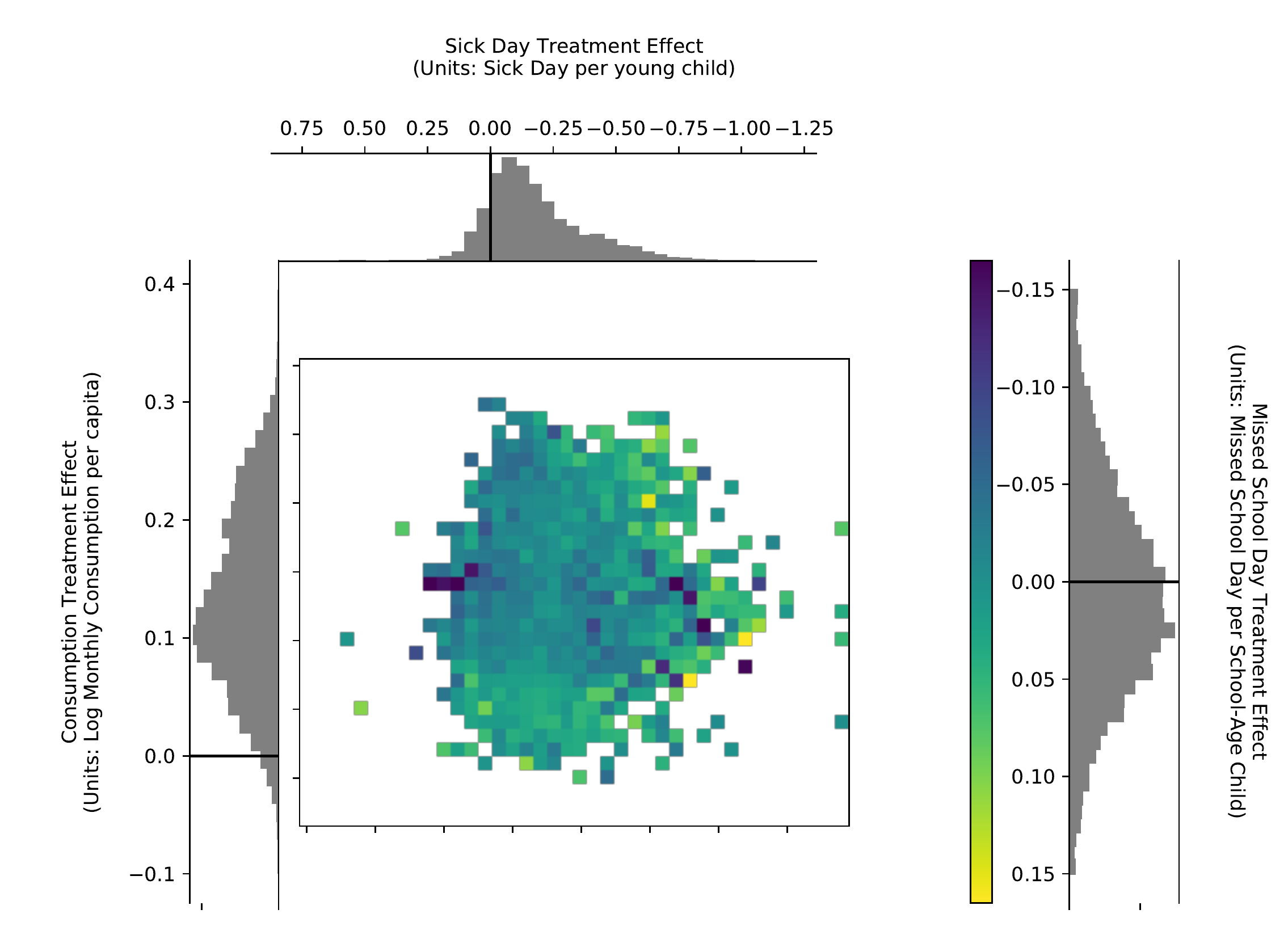}
\end{center}
 \begin{minipage}{1\textwidth}
    \footnotesize
    \emph{Notes:} Heterogeneous treatment effects of PROGRESA, estimated using causal forests \citet{wager_estimation_2018}. Histograms show marginal treatment effects on log Consumption (left), Health (top), and School Attendance (right). Center figure shows joint distribution, where each cell corresponds to a combination of consumption and health treatment effects, and is colored according to average treatment effect on attendance. Households without at least one young and one school-age child are omitted from the figure.
 \end{minipage}
\end{figure}

However, these treatment effects were heterogeneous. We estimate this heterogeneity $\Delta\hat{v}_{j}(\tilde{\mathbf{x}}_{i})$ using \citeauthor{wager_estimation_2018}'s \citeyearpar{wager_estimation_2018} causal forest method, which
recovers heterogeneous treatment effects nonparametrically, and includes restrictions to limit overfitting. %This approach allows for more flexible and precise estimates of heterogeneity than possible with linear methods \citep{wager_estimation_2018}.
 Figure~\ref{fig:Distribution-of-Estimated-CF} shows that different households benefit by different amounts across the three outcomes. In particular, the program increased the consumption of indigenous households more than non-indigenous households. This can be seen in the fact that indigenous status is the most important feature in the causal forest (Appendix Table~\ref{tab:Treatment-Effect-Coefficient-1}, column 1). The heterogeneity by indigenous status is also evident in Appendix Figure \ref{fig:Binscatter-CF}, which shows residualized treatment effects, estimated after removing variation explained by the other covariates.

While our main analysis relies on causal forests, which allow for more flexible and precise estimates of heterogeneity than linear models, the approach described in Section~\ref{sec:estimation} can be used with alternative methods for estimating heterogeneous treatment effects. Corresponding results for OLS are reported in Appendix Section \ref{subsec:OLS-Treatment-Effect} and Appendix~Figure~\ref{fig:Binscatter-OLS}.%
\footnote{Comparing the causal forest estimates in Appendix~Figure~\ref{fig:Binscatter-CF} with the OLS estimates in Appendix~Figure~\ref{fig:Binscatter-OLS}, the relative flexibility of causal forests is apparent. While the general pattern of heterogeneity is often consistent across both  methods, the causal forest method better captures non-linearity.}

\subsubsection{Implied Policy Preferences}
\label{sec:implied_prefs}

Next, given that we predict the policy would have impacts $\Delta \hat{v}_{ij}$ on household $i$, we use our method to back out the implied preferences consistent with ranking that household at position $z_i$. Table~\ref{tab:Allocation-Rule-Actual} column 2 reports the preferences that are consistent with the ranking $\boldsymbol{z}$. 
The first block of rows shows the implied welfare weights ($\boldsymbol{\omega}$), and the second block shows implied impact weights ($\boldsymbol{\lambda}$ and $C$) and the standard deviation of the error term ($\sigma$).

% Although it is not very informative to directly compare magnitudes between ordinal logit and our estimated welfare weights (ordinal logit models are isomorphic under multiplication by a positive number), we can compare ratios of coefficients, and treatment effects. 

We find that when the differential benefit that indigenous households face is accounted for, the decision rule actually implicitly places \emph{lower} welfare weight on indigenous households (by $11.7\%=1.01^{-12.4}$). Likewise, part of the negative weight on higher-income households can be explained by slightly lower marginal treatment effects for consumption for those households, and so the model infers moderately less negative implicit welfare weight on income when taking these heterogeneous treatment effects into account.

% Overall, we find that allocations are consistent with welfare weights that rank households 14.2 percentiles lower if indigenous, 12.5 percentiles lower for each standard deviation increase in household income, 15.2 percentiles higher for each standard deviation increase in household size, 20.4 percentiles lower for each standard deviation increase in age of the household head, and 12.5 percentiles lower for every standard deviation increase in the education of the household head.\footnote{We compute the average percentile change by first computing how each
% household's projected ranking would shift, given different covariates,
% and then taking the median change over all households.} 

Our estimates of weights on different impacts are imprecise, so we focus on the bounds implied by 95\% confidence intervals. These suggest that the Mexican government's initial
allocation rule is consistent with valuing each day of school attendance at less than 36\% of daily per capita consumption (the point estimate suggests a positive value of 3\%). They also suggest that the rule is consistent with valuing each prevented sick day per young child at less than 2\% of daily per capita consumption (the point estimate actually suggests a negative value of 8\%). The bounds for the value of schooling cover estimates of the returns to schooling from the literature; based on a review of multiple studies, \citet{psacharopoulos_returns_2018} suggest a 9\% average return to a year of schooling.

Most of the implied value of the program comes from its impact on consumption, followed by the effect of providing the program independent from its effect on measured outcomes (the constant
term $C$). The value of $C$ of 0.47 log points corresponds with an average of 179.4 pesos of consumption per person per month. The fact that this is larger than the average transfer of 33.9 pesos per person per month \citep{john_hoddinott_impact_2004} suggests that the policy may implicitly value a peso of recipient consumption less than a peso of transfer. (Our estimates denominated in pesos are relative to the value placed on consumption gains.)

\begin{table}
\centering \caption{Assessing Decision Rules\label{tab:Allocation-Rule-Comparison}}
\renewcommand{\arraystretch}{1.3}
\begin{threeparttable}
\begin{footnotesize}
\begin{tabular}{lccccc}
\toprule
 & & (1) & (2) & & (3) \\
 &  & \multicolumn{2}{c}{Implied Preferences (Estimated)} &  & Stated Preferences\\
 &  & (1999 Pov. Score)  & (2003 Pov. Score)  &   & (Resident survey)\\[.2em]
 \cline{3-4} \cline{6-6}
\\[-.5em]
\multicolumn{5}{l}{\textbf{Welfare Weights $log_{1.01}(\boldsymbol{\omega})$} \emph{(number of 1\% increments)}}\\
{Indigenous}  &  & {-12.4 (4.2)}  & {1.5 (3.0)}  &  & {-6.1 (6.4)}\\
{log(Income)}  &  & {-14.3 (5.4)}  & {-1.8 (1.4)}  &  & {-20.2 (8.5)}\\
{Household Size}  &  & {5.6 (2.1)}  & {1.9 (1.3)}  &  & {1.6 (2.0)}\\
{Head Age}  &  & {-1.0 (0.6)}  & {-0.03 (0.07)}  &  & {0.4 (0.3)}\\
{Education (HS or higher)}  &  & {-39.9 (21.5)}  & {-9.8 (7.8) }  &  & {-6.3 (3.6)}\\ [.5em]
\multicolumn{6}{l}{{\textbf{Impact Weights}{
}\emph{(log points of daily consumption)}{}}}\\
{Missed Schooling (per day)}  & \textbf{$\lambda_{1}$}{
}  & {-0.03 (0.17)}  & {0.02 (0.2)}  &  & {-0.35 (0.15)}\\
{Sickness (per child sick day)}  & \textbf{$\lambda_{2}$}{
}  & {0.08 (0.05)}  & {0.16 (0.12)}  &  & {-0.34 (0.15)}\\
{Value Regardless of Impact}  & {$C$}  & {0.47 (3.75)}  & {2.82 (12.52)}  &  & {.}\\[.4em]
{$\sigma$}  &  & {0.17 (0.17)}  & {0.31 (0.28)}  &  & {.}\\[.4em]
{N}  &  & {6642}  & {6642}  &  & {310}\\
    \bottomrule \\[-2.5ex] 
\end{tabular} 
\end{footnotesize}
\begin{tablenotes}[normal,flushleft]
\footnotesize
\item \emph{Notes}:
Columns 1-2 are estimated using our method, using causal forests to estimate heterogeneous treatment effects. Column 1 estimates model using 1999 poverty scores; column 2 using 2003 poverty scores. Column 3 indicates stated preferences based on a survey of Mexican residents. Standard errors in columns 1-2 are computed using a two-step bootstrap procedure that accounts for uncertainty in both treatment effects and preference parameters. 
% To calculate standard errors in columns 1-2, observations are drawn with replacement before estimation of the treatment effects and the welfare and impact weights. Treatment effects are then estimated from these bootstrapped samples, and welfare and impact weights estimated from these bootstrapped treatment effect estimates; the standard errors reported are the standard deviation across bootstrapped welfare and impact weight estimates. 
We exclude bootstrap draws (0 draws for the 1999 ranking, 1 draw for the 2003 ranking, out of 50 for each) that converged to corner solutions against the zero lower bound for omega. Standard errors in column 3 are computed directly from survey responses.
\end{tablenotes}
\end{threeparttable} 
\end{table}

%%%%%%%%%%%%%%%%%%%%%%%%%%%%%%%%%%%%%%%%%%%%%%%

\subsubsection{Alternative Preferences}

Our framework also makes it possible to  compare the preferences consistent with alternative policies. For instance, in the PROGRESA case, the Mexican government expanded the program in 2003, using a different poverty score to increase the priority of older and smaller households \citep{targeting_densification}. As shown in column 2 of Table~\ref{tab:Allocation-Rule-Comparison}, our method reveals that this new poverty score implicitly placed more welfare weight on richer households, and slightly less weight on larger and younger households. The impact weights are also imprecisely estimated; the 95\% confidence intervals suggest the valuation of a missed day of school below 36\% of consumption and of a young child sick day below 7\% of consumption.

Table~\ref{tab:Allocation-Rule-Comparison} also illustrates how the implemented policy (column 1) compares to the median stated preferences of residents, as reported in the survey we conducted in 2021 (column 3).  The welfare weights implied by the implemented policy are similar to resident preferences: we fail to reject differences in all but age of the household head (which is small in magnitude). On average, survey respondents value impacts on children more: sick days at 34\% of daily consumption and school attendance at 35\% of daily consumption, though both of these estimates are imprecise.

\subsection{Counterfactuals}

We next consider the reverse problem: given preferences, what would the resulting policy look like? In the PROGRESA example, Table~\ref{tab:Allocation-Rules} compares the policy's true allocation (column 1) to counterfactual allocations that would have resulted from alternative preferences (columns 2-6). Panel A indicates which preferences are used. We allow the welfare weights to be those estimated from the 1998 policy (columns 1, 4-6), those elicited from the resident survey (column 2), or fixed to weight all households equally (column 3).  We allow the impact weights to be those estimated from the 1998 policy (columns 1 and 3), those elicited from the resident survey (column 2), or to only value one outcome (columns 4-6).
Panel B indicates the decision rule implied by those preferences. Panel C shows the average outcomes that would be expected under the hypothetical policy, assuming it treated the same number of households as the implemented policy.

%%%%%%%%%%%%%%%%%% TABLE  %%%%%%%%%%%%%%%%%%%%%%%%%%%%%%%%%%%%
\begin{landscape} 
\begin{table}
\centering
\caption{Designing Decision Rules\label{tab:Allocation-Rules}}
\renewcommand{\arraystretch}{1.2}
\begin{threeparttable}
\begin{footnotesize}
\begin{tabular}{lccccccc}
\toprule 
 &(1) &  (2)  & (3) &  & (4) & (5) & (6) \\
 
 % & \multicolumn{5}{c}{\textbf{Actual}} &  \multicolumn{4}{c}{\textbf{Counterfactual}}\\[.5ex] 
&HH Poverty &  Resident &  Equal Welfare &  & \multicolumn{3}{c}{Policy only values impact on:} \\
 & {Score } &  \multicolumn{1}{c}{{Preferences}} &  \multicolumn{1}{c}{{Weights}} &  & {Education } & {Health } & {Consumption } \\
 \cline{2-4} \cline{6-8} 
 \\[-.5em]
 \textit{Panel A: Preferences}   &  &  &  &  &  &  &  \\
{Welfare Weights $\boldsymbol{\omega}$  } & {Estimated} &  From survey  & {Unity } &  & {Estimated} & {Estimated} & {Estimated} \\
{Impact Weights $\boldsymbol{\lambda}$  } & {Estimated} &  From survey   & {Estimated} &  & {Only education } & {Only health } & {Only consumption } \\ [.5em]
 \multicolumn{8}{l}{\textit{Panel B: Implied decision rule (priority over covariates, in 1\% increments)}}  \\
{$\;\;$ Indigenous } & {38.6} &   {17.2}   & {143.3} &  & {-48.2} & {53.0} & {176.2} \\
{$\;\;$ log(Income) } & {-24.9} &  {-88.4}  & {-10.0} &  & {89.8} & {-28.5} & {-26.4} \\
{$\;\;$ Household Size } & {13.0 } &  {15.0}  & {2.31} &  & {-13.8} & {4.3} & {8.9} \\
{$\;\;$ Head Age } & {-2.3 } &  {3.3}  & {-0.6} &  & {-10.8} & {-1.1} & {-1.6} \\
{$\;\;$ Education } & {-240.2} & {32.4}  & {12.7} &  & {24.2} & {-93.9} & {-49.8} \\ [.5em]
\multicolumn{8}{l}{\textit{Panel C: Counterfactual outcomes (monthly)}}  \\
{{$\;\;$ Log Consumption (pesos)}  } & {{4.852} } &  {{4.853}  }  & {{4.875}  } &  & {{4.849}  } & {{4.843}  } & {{4.874}  }  \\
{{$\;\;$ Missed school (days/child)}  } & {{0.168}  } &  {{0.167}  } &  {{0.164}  } &  & {{0.140}  } & {{0.171}  } & 0.165\\
{{$\;\;$ Sickness (sick days/child)}  } & {{0.637}  } & {{0.609} }  & {{0.655}  } &  & {{0.647} } & {{0.567}  } & {{0.635}  }\\[.4em]
{$N$  } & {{6642}  } &  {{6642}  }  & {{6642}  } &  & {{6642}  } & {{6642}  } & {{6642}  } \\
\bottomrule
\end{tabular} 
\end{footnotesize}
\begin{tablenotes}[normal,flushleft]
\footnotesize
\item \emph{Notes}: 
Table shows the distributional and outcome effects of designing decision rules using our framework. Panel A indicates which weights are used to prioritize households. Column 1 uses the ranking assigned by PROGRESA. Column 2 uses preferences elicited in a survey we conducted of Mexican residents. Column 3 projects the ranking as though the policy did not prioritize certain types of households, and was based on preferences over outcomes estimated in Table~\ref{tab:Allocation-Rule-Comparison}. Columns 4-6 indicate what would have happened if the policy used the estimated weights over households but only valued about impacts on education/health/consumption. Panel B shows the distributional effects of each column's preferences, by estimating the implied priority ranking across households. Panel C shows each policy's expected average outcomes, calculated using estimates of heterogeneous treatment effects.
\end{tablenotes}
\end{threeparttable}
\end{table}
\end{landscape}

%%%%%%%%%%%%%%%%%%%%%%%%%%%%%%%%%%%%%%%%%%%%%%%%%%%%%%

\paragraph{Survey-based estimates of resident preferences}

Column 2 shows the allocation that would result from imposing the preferences of residents as revealed by the survey. Relative to the actual policy in column 1, the hypothetical policy in column 2 reduces the prioritization of indigenous households, and increases the prioritization of  poor households. Other household attributes are similarly prioritized under the two policies. In Panel C, we see that the policy consistent with resident preferences would slightly reduce child sickness relative to the implemented policy.

\paragraph{Alternate welfare weights}

When welfare weights are set equal across households (column 3), the
resulting score prioritizes indigenous households by a much larger factor, and lowers the priority given to lower income and larger households. 
% When welfare weights rank households solely by log income (column 3), the resulting score deprioritizes households with more small children and also deprioritizes indigenous households.

% \subsubsection{Technocratic impact weights}

% In columns 4-5 of Table \ref{tab:Allocation-Rules}, we keep the original
% welfare weights but assume technocratic impact weights, as might be
% input from external valuations. We do not intend to take a stand on
% these valuations, so these results should be viewed as speculative.
% We demonstrate results assuming valuations of 1000 pesos (\$100) per
% DALY and 2143.35 pesos per missed school day. The $\boldsymbol{z}'$
% ranking implied by our assumed weights is quite similar to the original.
% By contrast, changing the household covariate welfare weights to an
% equal weighting across households so that $\mu(\mathbf{x}_{i})\equiv1$
% leads to positive weight on income and on household number of adults.

\paragraph{Prioritizing specific welfare outcomes}

While in practice implemented policies may balance multiple outcomes,
in columns 4-6 of Table \ref{tab:Allocation-Rules}, we present counterfactual allocations that would result in the extreme case where a policy was designed to improve only a single outcome. For instance, a policy designed to maximize education would prioritize smaller households and those with \emph{higher} income (column 4). On the other hand, if only health effects were valued, the policy would slightly increase the prioritization of indigenous households (column 5). Finally, a policy that maximized consumption with no explicit consideration of health or education (column 6) would place much greater priority on households where the head is indigenous and reduce the penalty on education. 

\paragraph*{}

\begin{figure}[t]
\begin{centering}
\caption{Expected Program Impacts under Alternative Preferences}\label{fig:allocations3d}
    
    \includegraphics[width=0.6\columnwidth]{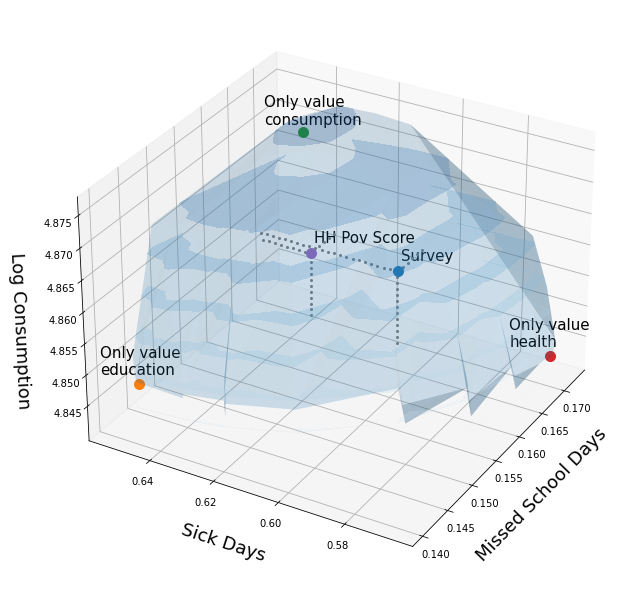}

\end{centering}
    {\footnotesize \textit{Notes:} Figure shows the frontier of possible average welfare impacts that would have resulted from different allocations of PROGRESA. Each axis indicates the expected average impacts for a given welfare outcome. Labeled points correspond to particular allocations described in Table \ref{tab:Allocation-Rules}. Appendix Figure~\ref{fig:allocations3d_survey} shows the frontier when outcomes are scaled by welfare weights.}
\end{figure}

Understanding the policies that would result from extreme preferences can help in understanding the full set of potential policies, and what those policies imply. In the PROGRESA case, Figure~\ref{fig:allocations3d} characterizes the frontier of possible average welfare impacts that would result from different allocations of the program. This frontier is shown as a convex hull with contour lines; the labeled points correspond to the policies given in the columns of Table~\ref{tab:Allocation-Rules}. Policies that only value a single outcome lie at the corners of the outcome space. The implemented program (`HH Poverty Score') is close to the allocation consistent with the survey of Mexican residents preferences; neither are quite on the frontier with respect to unweighted outcomes, but both are close. (They are on the frontier of outcome spaces scaled by the corresponding welfare weights, see Appendix Figure~\ref{fig:allocations3d_survey}.) More broadly, this method makes it possible to navigate program design in outcome space, rather than implementation space.

%Panel~(a) shows the average outcomes, and Panel~(b) the outcomes multiplied by the welfare weights implied by the PROGRESA poverty score, derived using our method.%
%\footnote{Appendix Figure~\ref{fig:allocations3d_survey} shows the outcomes multiplied by the welfare weights implied by our survey respondents.} %
% The frontier of possible outcomes is shown as a convex hull with contour lines. The implemented program (`baseline') achieves outcomes that are not on the frontier according to raw outcomes (a), but are on the frontier of outcomes weighted by its implied welfare weights (b). Policymakers who only valued a single outcome would land at a corner of the outcome space. The allocation consistent with the survey of Mexican residents preferences lies close to the implemented allocation. The method makes it possible to navigate program design in outcome space (a) or welfare space (b), rather than implementation space (the decision rules shown in Table~\ref{tab:Allocation-Rules} Panel B).

\subsection{Extensions}

\paragraph{If only an allocation is observed}

In many settings, priority rankings are not available. Our method can still be used when the analyst observes only the final allocation (e.g., who receives the program, or who is admitted). This is because the binary indicator of whether a household received an allocation represents a (short) ranking. In the PROGRESA example, Table~\ref{tab:Binary-Ranking} demonstrates that when our method is applied to a binary allocation ($z(\mathbf{x}_i)=1\{i \textnormal{ above median rank}\}$), point estimates are similar to those reported in Table~\ref{tab:Allocation-Rule-Actual}. Though the estimates are much less precise under the binary allocation, the qualitative insights are the same.

\paragraph{Testing models of welfare}
The method can also be used to test whether policies are internally consistent with a postulated welfare function. If there is more than one potential treatment or policy, one could test the hypothesis that they apply the same welfare weights for each one. If that hypothesis is rejected, one can rule out that the policies are consistent with utilitarianism, given ex post information.

% \paragraph{Bounding}

% In settings where heterogeneous treatment effects are unavailable
% or difficult to measure, one could alternately use this framework
% to bound the treatment effects that would be consistent with stated
% preferences.

\section{Conclusion}

Policy discussions commonly revolve around the mechanics of implementation, rather than more fundamental notions of utility and welfare weights. This paper demonstrates how these discussions can be inverted. We provide a method to recover the primitives consistent with observed policies, using a model of preferences in conjunction with modern  methods for estimating heterogeneous treatment effects, and demonstrate how to convert between welfare and allocation space.

We develop this approach and apply it to a large anti-poverty program in Mexico, to estimate the preferences consistent with the program's implementation. This analysis reveals that, after accounting for heterogeneity in treatment effects, the program's allocation placed higher weight on the welfare of poor and large families, and lower weight on indigenous households. The implied value of each missed school day and child sick day is estimated imprecisely but our confidence intervals do not rule out valuations estimated in prior work. 

This framework could be used in several ways. To begin, it could be used
to characterize the realized allocations of an existing program, to provide an indication of the preferences they imply. This, in turn, can provide a way to audit an existing program, to help hold policymakers accountable for past decisions
\textendash{} and in particular, to evaluate whether the implemented
allocation reflects the stated goals of the policy. Perhaps most importantly,
this approach can be used to adjust existing policies to better align with those goals.

% In particular, it can demonstrate how different values over welfare outcomes and population subgroups would produce different allocations, and quantifies the welfare impacts of these adjustments.

%%%%%%%%%%%%%%%%%%%%%%%%%%
%  BIBLIOGRAPHY
%%%%%%%%%%%%%%%%%%%%%%%%%%
\clearpage
\begin{spacing}{1}
\bibliographystyle{aer}
\bibliography{Manipulation}
\end{spacing}

%%%%%%%%%%%%%%%%%%%%%%%%%%
%  APPENDICES
%%%%%%%%%%%%%%%%%%%%%%%%%%
\clearpage
\newpage

\section*{\center{Appendices}}
\renewcommand{\thetable}{A\arabic{table}}
\renewcommand{\thefigure}{A\arabic{figure}}
\renewcommand{\thesection}{A\arabic{section}}
\setcounter{table}{0}
\setcounter{figure}{0}
\setcounter{section}{0}

\section{Generalized curvature in utility components\label{subsec:generalizedcurvature}}

This section considers what will be measured if the utility functions are assumed to be linear ($\tilde{g}_j(y)=y$) but in fact the true utility functions $g_j(y)$ have curvature.
The true impact of the program on component of utility $j$ is then:

\[
\Delta v_{ij} = g_{j}(y_{ij}^{1}) - g_{j}(y_{ij}^{0})
\]

Taking a Taylor approximation from the factual level $y_{ij}$, we have $g_j(y_{ij}+\delta)\approx g_j(y_{ij})+\delta\cdot g_j'(y_{ij})$. Thus for any $g_j(\cdot)$ we have:

\[
\Delta v_{ij} \approx g_{j}(y_{ij}) - g_{j}(y_{ij}) + \Delta y_{j}(\tilde{\mathbf{x}}_i)\cdot g_j'(y_{ij}) = \Delta y_{j}(\tilde{\mathbf{x}}_i) \cdot g_j'(y_{ij})
\]

We can then express the utility benefit of treating $i$ as:

\[
\Delta S_{i}\approx\underbrace{\mu(\mathbf{x}_{i})g'_0(y_{i0})}_{\tilde{\mu}(\mathbf{x}_{i},\{y_{i0}\})}\left[\hat{\Delta y_{0}}(\tilde{\mathbf{x}}_{i}) + \sum_{j} \underbrace{\lambda_j(\mathbf{x}_{i})g'_j(y_{ij})}_{\tilde{\lambda}_j(\mathbf{x}_{i},\{y_{ij}\})} \hat{\Delta y_{j}}(\tilde{\mathbf{x}}_{i}) \right]
\]

% \[
% \Delta S_{i}\approx\underbrace{\mu(\mathbf{x}_{i})u'(y_{i0})}_{\tilde{\mu}(\mathbf{x}_{i},y_{i0})}\cdot\left(\hat{\Delta g_{0}}(\mathbf{x}_{i})\right)
% \]

This implies that if we do not specifically account for curvature and
estimate a linear model, the welfare and impact weights we estimate ($\tilde{\mu}$ and $\tilde{\boldsymbol\lambda}$)
are approximately a combination of the underlying welfare and impact weights
($\mu$ and $\boldsymbol\lambda$) and any curvature in the utility functions ($g_j'$),
as long as the baseline value of the outcome ($y_{ij}$) is included as a characteristic along which these weights can vary ($\tilde{\mathbf{x}}_i$). If the true utility is linear, then
$\tilde\mu$ coincides with ${\mu}$ and $\tilde{\boldsymbol\lambda}$ with $\boldsymbol\lambda$. Otherwise, utility curvature
multiplies the weights.

% \subsubsection{Multiple Outcomes}

% If we allow curvature in all dimensions but keep our initial functional
% form that requires all dimensions to diminish by the same factor (eg
% baseline income):

% \[
% \Delta S_{i}\approx\underbrace{\mu(\mathbf{x}_{i})u'_{?}(y_{i?})}_{\tilde{\mu}(\mathbf{x}_{i},\{y_{ij}\})}\left[\hat{\Delta g_{0}}(\mathbf{x}_{i})+\sum_{j}\lambda_{j}\hat{\Delta g_{j}}(\mathbf{x}_{i})\right]
% \]

% That doesn't map to a standard utility function.

% We can instead estimate an outcome specific $\tilde{\mu_{j}}$ which
% combines welfare weight and curvature:

% \[
% \Delta S_{i}\approx\sum_{j}\underbrace{\mu(\mathbf{x}_{i})u'_{j}(y_{ij})}_{\tilde{\mu_{j}}(\mathbf{x}_{i},y_{ij})}\cdot\hat{\Delta g_{j}}(\mathbf{x}_{i})
% \]

% This is more of a nonparametric approach. After recovering $\tilde{\mu_{j}}$
% we can compare between outcomes, or to interpret, impose a utility
% function form.

\pagebreak
\section{Data Cleaning Process}

The data for the evaluation of PROGRESA is composed of household survey
responses from a sample of 506 villages from seven states across multiple
years. Three different survey years are used: a baseline survey in
October 1997, and two follow-up surveys in October 1998 and November
1999. Villages were randomly assigned to a treatment group which received the program at the beginning, and a control groups, which received it two years later.
Within villages in the treatment group, a poverty index score is computed
based on household income and assets, and all households meeting the
score requirement are eligible to receive the program's conditional
transfers. 
% In our specified timeframe, there were two rounds of eligibility
% classification: the first during the 1997 baseline survey and the
% second during a July 1999 survey. The second round increased the scope
% of the program to include more households, and we use this criteria
% to classify eligible households.

We compute a measure of average household monthly consumption per
member based on the survey responses. The October 1998 and November
1999 surveys ask households about the quantity consumed, quantity
purchased and amount of money expended on 36 common food items, as
well as expenditure for several non-food categories (in weekly/monthly/semi-annual
amounts). We use the information regarding quantity purchased and
amount of money expended to construct household-specific prices which
are then multiplied by quantity consumed (this helps to account for
the fact that households consume food that is self-produced in addition
to bought). If household-specific information is missing, we use locality,
municipality or state average prices (the smallest level available).

\pagebreak
\section{Preference survey\label{subsec:preference-survey}}

We additionally survey Mexican residents to elicit their preferences for different allocations of social welfare programs. We solicited responses to a survey from a nationally representative sample of computer users in Mexico, through a Qualtrics survey panel.

\subsection{Survey design}

\begin{figure}
\caption{Welfare Weight Survey Question Example\label{fig:Survey_WelfareWeight}}
\begin{centering}
\includegraphics[scale=0.4]{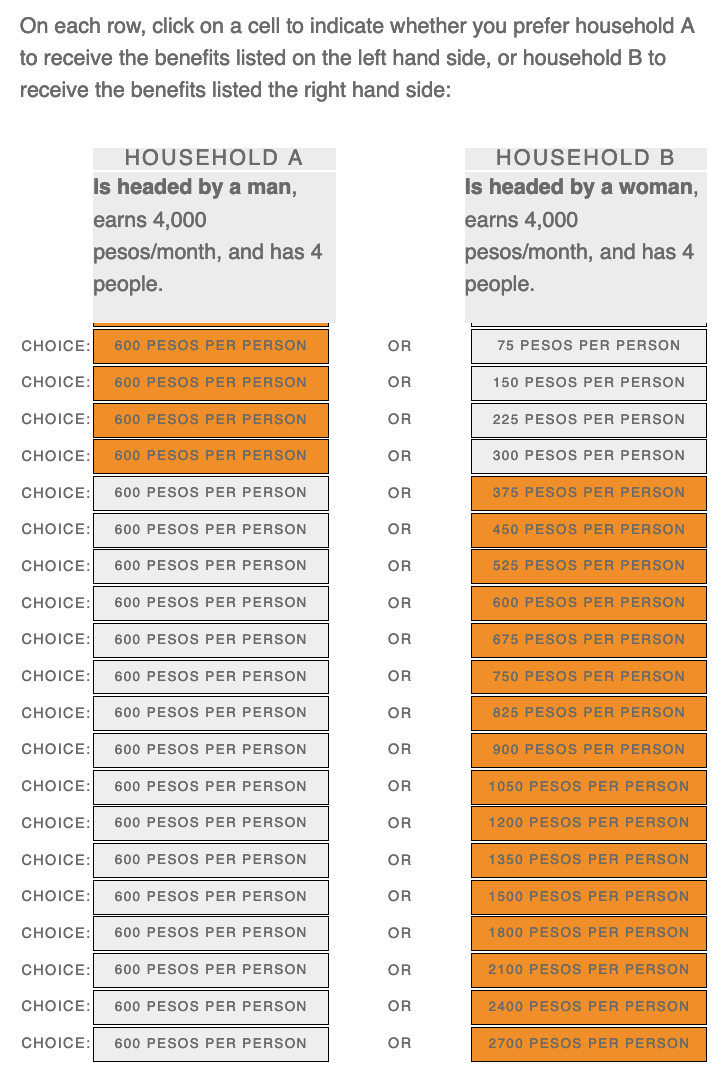}
\par\end{centering}
\smallskip
\footnotesize{\textit{Notes:} Respondents saw a version of this question translated into Spanish.}
\end{figure}

\begin{figure}
\caption{Impact Weight Survey Question Example\label{fig:Survey_ImpactWeight}}
\begin{centering}
\includegraphics[scale=0.4]{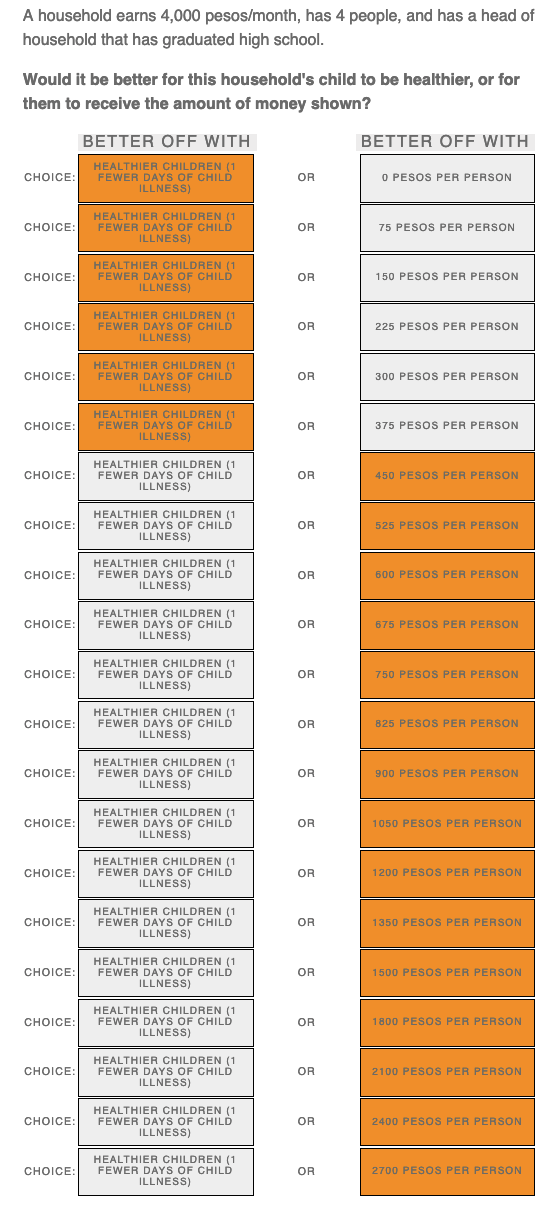}
\par\end{centering}
\smallskip
\footnotesize{\textit{Notes:} Respondents saw a version of this question translated into Spanish.}
\end{figure}

After obtaining consent and an initial information screen, participants were asked their preferences for allocating benefits to different types of households. The survey was translated in Mexican Spanish. First, respondents were asked to select which attributes the government should consider when prioritizing which households receive cash transfers, from a list (age, income, household size, education, agricultural, indigenous, and gender). Second, subjects were asked to make monetary allocation decisions between different households using multiple price lists (see Figure~\ref{fig:Survey_WelfareWeight} for an example). In each, one focal attribute differed between the households, and two other control attributes were held fixed. We randomized which controls were included, the order they were presented, and the scale of the tradeoff.\footnote{Each participant saw base tradeoff numbers multiplied by 1x, 2x, or 3x, selected at random.} Each subject filled in one price list for each focal attribute. Third, for a particular household, subjects were asked to make allocation decisions between money and education and child health using multiple price lists (see Figure~\ref{fig:Survey_ImpactWeight}). The description of the household included three randomly selected control attributes. Finally, subjects were asked for basic demographics.

\subsection{Estimation}

We use the survey responses to estimate $\boldsymbol\omega$ and $\boldsymbol\lambda$:

% Si=[1+xi][g0(xi)+j>0jgj(xi)+C]

To  identify $\boldsymbol\omega$, compare impacts in dollars of consumption (where other impacts $\Delta g_j(x_i)=0$). If individual $i$ differs from $i'$ only in attribute $j$ and the crossover point is $\Delta g_0(x_i)=a$ and $\Delta g_0(x_{i'})=b$, then
% For additive welfare weights:
% \begin{align*}
% [1+\omega_j x_{i,j}]a &= [1+\omega_j x_{i',j}]b\\
% \omega_j &= -\frac{b-a}{x_{i',j}b-x_{i,j}a}    
% \end{align*}
% For multiplicative welfare weights:
\begin{align*}
\omega_{-j}^{x_{i,-j}}\omega_j^{x_{i,j}}a &= \omega_{-j}^{x_{i',-j}}\omega_j^{x_{i',j}}b \\
\omega_j &= \left(\frac{b}{a}\right)^{\frac{1}{x_{i,j}-x_{i',j}}}
\end{align*}

To identify $\boldsymbol\lambda$, now instead hold fixed individual attributes, and consider impacts on different outcomes. If the crossover point is $\Delta g_0(x_i)=a$ and $\Delta g_j(x_i)=b$ then $\lambda_j=\frac{a}{b}$.

\subsection{Validation}

The design included several checks to ensure that respondents took the survey seriously. First, prior to the survey, participants were asked, `We care about the quality of our survey data and hope to receive the most accurate measure of your opinions, so it is important to us that you thoughtfully provide your best answer to each question in the survey. Do you commit to providing your thoughtful and honest answers to the questions in this survey?' Only participants who answered `I will provide my best answers' were invited to continue with the survey. Second, after reading the instructions, participants responded to five simple questions to validate understanding of the study. In order to complete the study, participants had to respond correctly. Third, the survey included controls to ensure that participants spent adequate time on each question. The submit button for the main exercises appeared only after a 5 second delay.\footnote{The implementation of this in Qualtrics made it possible for participants to advance if this time had elapsed, even if a multiple price list question had not been answered. For this reason, a handful of participants did not respond to all questions.} Additionally, participants who were completing the survey too quickly (less than half the median elapsed time in the pilot survey) were removed from our sample, following a standard quality protocol used by Qualtrics. Fourth, in the final demographic survey, respondents were asked to rate the following three statements along the same Likert scale ranging from ‘Strongly Disagree’ to ‘Strongly Agree’: ‘I made each decision in this study carefully’, ‘I made decisions in this study randomly’, and ‘I understood what my decisions meant.’ A careful respondent should agree with the first and last statement but disagree with the middle; agreement or disagreement with all statements reveals that a respondent made careless decisions. We restrict the sample to only respondents who disagreed that they had made decisions randomly.\footnote{Apart from two pilot respondents.} 91\% of respondents agreed with the first and last statement, and disagreed with the middle; 58\% did so strongly.

There was an optional comment box at the conclusion of the survey; 49\% of respondents filled in a comment, suggesting a high level of engagement with the survey. Although some respondents used the box to indicate some confusion with the selector interface, several respondent affirmatively to the approach of basing policy on resident preferences, such as (translated to English):

\begin{itemize}
    \item `Excellent that they do these surveys to assess the policies of support to families'
    \item `I think this survey was very important since the benefits that sometimes come are the same for all people and the situations of each person are not considered. For some it may be enough but for others it is too little.'
    \item `excellent survey, hopefully and we could society decide these support, because that is how we would eradicate poverty'
\end{itemize}

\pagebreak
\section{OLS Treatment Effect Estimates\label{subsec:OLS-Treatment-Effect}}

\subsection{Estimation: OLS}

One can also use linear regression to estimate heterogeneous treatment effects. We follow \citet{djebbari_heterogeneous_2008}, allowing treatment effects
to vary by age and gender composition of the household, total household
size, and several characteristics of the household head: education
level, indigenous status, gender, working in the agricultural sector,
and age.%
\footnote{We depart from \citet{djebbari_heterogeneous_2008} in that we omit poverty scores and village marginality index and their respective interactions in the list of covariates, to avoid potential correlated errors from using these rankings in both the treatment effect estimates and in the preference-learning method.} Formally, we estimate:

\begin{equation}
g_{ij}=\beta_{0}+\boldsymbol{\beta}_{\mathbf{x}}\mathbf{x}_{i}+(\beta_{T}+\boldsymbol{\beta}_{T\mathbf{x}}\mathbf{x}_{i})T_{i}+e_{i}\label{eq:impactOLS}
\end{equation}

where $g_{ij}$ is the endline outcome, $\mathbf{x}_{i}$ is the vector
of baseline covariates, and $T_{i}\in\{0,1\}$ is a dummy variable
for treatment status of household $i$. This model allows endline
outcomes to differ systematically according to household covariates,
and additionally allows the treatment effect of PROGRESA to differ
across households according to their covariates.

We construct our variables for treatment effects from the predicted
values from our estimated equation~\eqref{eq:impactOLS}, as

\[
\Delta\hat{g}_{j}(\mathbf{x}_{i})=\hat{\beta}_{T}+\hat{\boldsymbol{\beta}}_{T\mathbf{x}}\mathbf{x}_{i}
\]

\subsection{Results}

On average over our sample, using OLS PROGRESA increased household log
monthly consumption by 0.07, to have reduced the number of sick days
per child by 0.22, and slightly increased the number of school days missed per
child by 0.026 (very close to zero).

\begin{figure}
\caption{Distribution of Estimated Treatment Effects (OLS)\label{fig:Distribution-of-Estimated-OLS}}
\includegraphics[scale=0.65]{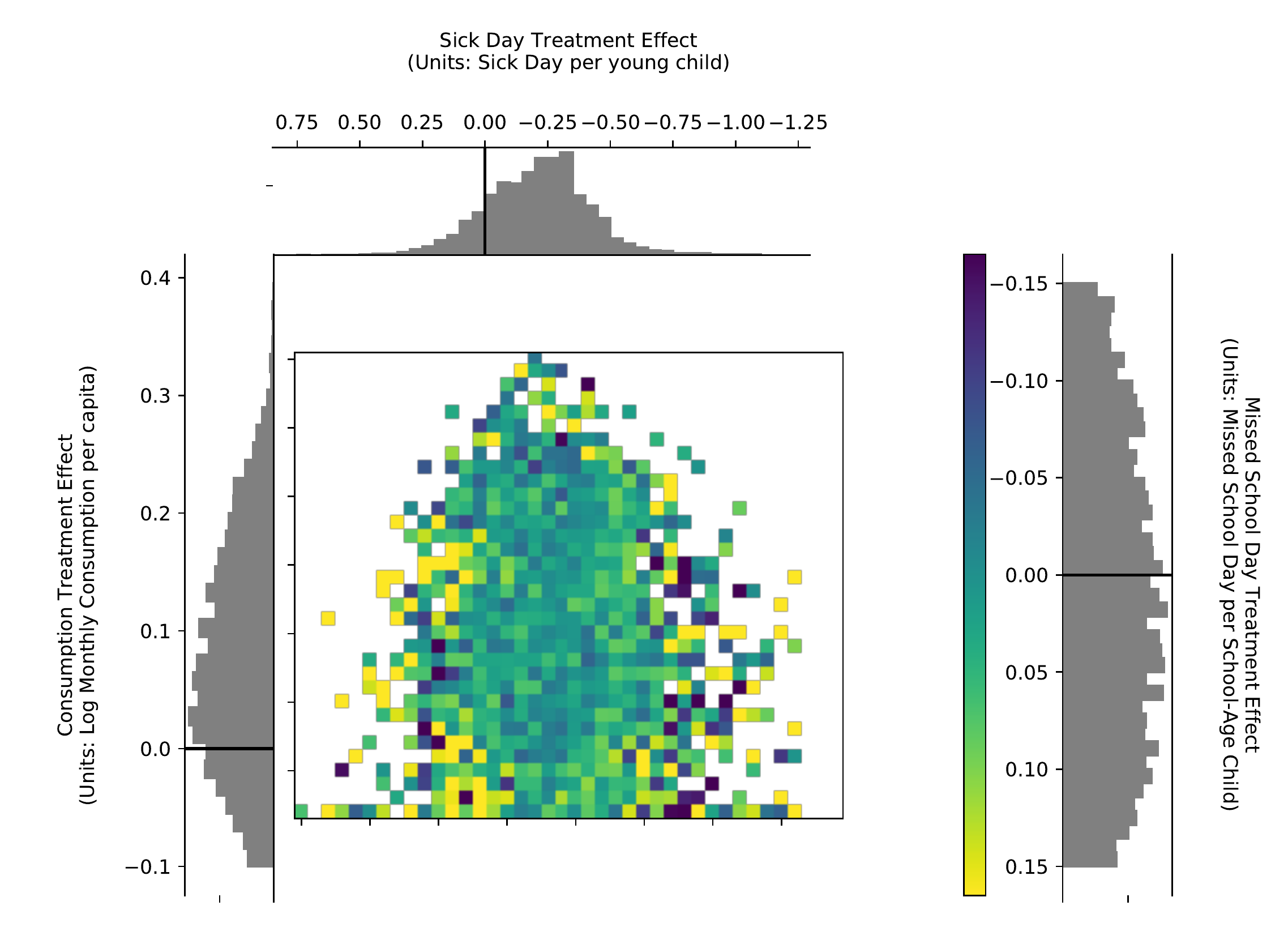}
\footnotesize{
 \textit{Notes:} Joint and marginal distributions of estimated treatment effects of PROGRESA conditional cash transfer on schooling, health, and consumption, estimated using OLS. Schooling treatment effects are measured over the number of missed school days per school-age child in a given household. Health treatment effects are measured over the number of sick days per young (0-5 years old) child in a given household. Consumption treatment effects are measured over per-person consumption in pesos in a given household. Marginal distributions for consumption and health treatment effects are shown over the y and x axes, respectively, and are binned together in the center figure. Average schooling treatment effects in each consumption-health-treatment-effect bin is shown by the fill color of the bin, according to the index of the legend on the right. The marginal distribution of schooling treatment effects is shown in parallel to this legend.
Note that missed school days and sick days are inferred to be "bads", according to our estimated weights, and so higher negative values for these treatment effects are associated with higher social utility. Note also that we drop households without children in the relevant age range for health and schooling treatment effects; the above graphs show only TEs for households for which these TEs are defined.}
\end{figure}

However, the effects of the program differ across households. The
overall distributions of treatment effects by outcome for OLS, are
presented in Figure \ref{fig:Distribution-of-Estimated-OLS}. The
distribution of estimated effects estimated under causal forest is
tighter, in particular for the schooling and health outcomes. With
OLS, we see a fairly strong correlation between health treatment effect
estimates and schooling treatment effect estimates, but with causal
forest this correlation is much less apparent. 

OLS coefficient estimates are presented in Table \ref{tab:Treatment-Effect-Coefficient-OLS},
with standard errors in parentheses. Similar to \citet{djebbari_heterogeneous_2008},
our OLS point estimates show that log consumption treatment impacts are
higher for households with indigenous status.\footnote{Note that our specification differs from \citet{djebbari_heterogeneous_2008}
in that we exclude the ranking metrics from the list of covariates.}

\clearpage
\pagebreak

\centering{\Large{\textbf{Appendix Exhibits}}}
\vspace{1cm}

%%%%%%%%%%%%%%%%%%%%%%%%%% TABLE %%%%%%%%%%%%%%%%%%%%%%
\begin{table}[hb]
\begin{threeparttable}
\caption{Descriptive Statistics\label{tab:Descriptive-Statistics}}
\global\long\def\arraystretch{1.1}%
\begin{tabular}{lc}
\toprule 
 & October 1998 mean \\ \hline
 \\[-.8em]
 Head of household:  & \tabularnewline
... Is indigenous  & 0.41\tabularnewline
... Age  & 41.13\tabularnewline
... Education (HS or higher)  & 0.005\tabularnewline
... Is male  & 0.94\tabularnewline
... Is an agricultural worker  & 0.65\tabularnewline
% Household poverty score (1997)  & 696.19 & 696.19\tabularnewline
% Village marginality index (1997)  & 0.47 & 0.47\tabularnewline
[1ex] Household size  & \tabularnewline
... Number of children less than 6 years old  & 1.97\tabularnewline
... Number of children 6-16 years old  & 2.81\tabularnewline
... Number of adults 17+ years old  & 2.54\tabularnewline
[1ex]Log monthly average per capita consumption (log pesos)  & 5.08 \tabularnewline
Average number of days a school-age child misses school  & 0.32\tabularnewline
Average number of days a young child is sick  & 1.07\tabularnewline
[1ex]Assigned to treatment group  & 0.61 \tabularnewline
\midrule 
 & \tabularnewline
[-2ex] N  & 6537\tabularnewline
\midrule 
\end{tabular}
\begin{tablenotes}[normal,flushleft]
\footnotesize
\item \emph{Notes}: Table shows the average levels in October 1998 of households matched to November 1999 survey sample. HS education defined as 12 years or more of education. Number of days a young child is sick, and number of days a school-age child misses school, are computed as an average over the number of children in the respective age group in the household. Sample restricted to only households with children in the targeted categories for health and schooling intervention (0-5 y.o., 6-16 y.o.) during the November 1999 survey.
\end{tablenotes}
\end{threeparttable}
\end{table}

%%%%%%%%%%%%%%%%%%%%%%%%%%  %%%%%%%%%%%%%%%%%%%%%%

%%%%%%%%%%%%%%%%%%%%%%%%%% TABLE %%%%%%%%%%%%%%%%%%%%%%
\begin{table}
\begin{threeparttable}
\caption{Feature Importance Estimates: Causal Forest\label{tab:Treatment-Effect-Coefficient-1}}
\global\long\def\arraystretch{1.2}%
\begin{tabular}{lccc}
\toprule 
 & \textbf{Log Consumption}  & \textbf{Schooling} & \textbf{Health} \tabularnewline
 & Monthly per capita  & \# days missed school & \# Sick days \tabularnewline
 & (pesos)  & per child & per child \tabularnewline
\midrule 
head age  & 0.112  & 0.308 & 0.228 \tabularnewline
household income 97  & 0.198  & 0.163 & 0.263 \tabularnewline
head indigenous  & 0.362  & 0.009 & 0.011 \tabularnewline
num child 3 to 5 yrs  & 0.01  & 0.072 & 0.162 \tabularnewline
num child less than 2 yrs  & 0.017  & 0.143 & 0.035 \tabularnewline
num adults  & 0.079  & 0.059 & 0.044 \tabularnewline
num child 6 to 10 yrs  & 0.074  & 0.02 & 0.044 \tabularnewline
num men at least 55 yrs  & 0.014  & 0.052 & 0.007 \tabularnewline
head agricultural worker  & 0.023  & 0.035 & 0.014 \tabularnewline
num women 20 to 34 yrs  & 0.009  & 0.026 & 0.037 \tabularnewline
num boys 11 to 14 yrs  & 0.011  & 0.03 & 0.023 \tabularnewline
num men 20 to 34 yrs  & 0.009  & 0.012 & 0.042 \tabularnewline
num girls 11 to 14 yrs  & 0.013  & 0.014 & 0.031 \tabularnewline
num girls 15 to 19 yrs  & 0.027  & 0.007 & 0.015 \tabularnewline
num boys 15 to 19 yrs  & 0.016  & 0.007 & 0.02 \tabularnewline
male head of household  & 0.005  & 0.023 & 0.004 \tabularnewline
num women 35 to 54 yrs  & 0.006  & 0.01 & 0.008 \tabularnewline
num men 35 to 54 yrs  & 0.008  & 0.006 & 0.007 \tabularnewline
num women at least 55 yrs  & 0.008  & 0.004 & 0.004 \tabularnewline
head education  & 0  & 0 & 0 \tabularnewline
\midrule 
{N}  & {6642}  & {6642} & {6642} \tabularnewline
\bottomrule
\end{tabular}
\begin{tablenotes}[normal,flushleft]
\footnotesize
\item \emph{Notes}: 
Feature importances as estimated from causal forest estimation of heterogeneous treatment impacts of PROGRESA on three outcome dimensions: log consumption (log monthly per capita consumption), schooling (number of missed school days per child), and health (number of sick days per child). Schooling and health sick days / missed school days measured over 28 days prior to survey. Estimates reflect 3 separate causal forest estimations for each respective outcome.
\end{tablenotes}
\end{threeparttable}
\end{table}

%%%%%%%%%%%%%%%%%%%%%%%%%%  %%%%%%%%%%%%%%%%%%%%%%

%%%%%%%%%%%%%%%%% TABLE  %%%%%%%%%%%%%%%%%%%%%%%%%%%%%%
\begin{table}
\centering \caption{If Only the Allocation is Observed\label{tab:Binary-Ranking}}
\renewcommand{\arraystretch}{1.2}
\begin{threeparttable} 
\begin{tabular}{llccc}
\toprule
 &  & \multicolumn{3}{c}{{\footnotesize{}{}Household Poverty Score}}\tabularnewline
& & \footnotesize{Observe full} &  & \footnotesize{Observe only}\\[-.5em]
&  & \footnotesize{ ranking} &  & \footnotesize{binary allocation}\\[.3em]
 \cline{2-5}
 \\[-1em]
&  & \multicolumn{3}{c}{\textbf{\footnotesize{}{}Log Welfare Weights $log_{1.01}(\boldsymbol{\omega})$}}\tabularnewline
 \footnotesize{Indigenous}  &  & {\footnotesize{}{}-12.4 (4.2)}  &  & {\footnotesize{}{}-9.71 (61.1)}\tabularnewline
{\footnotesize{}{}log(Income)}  &  & {\footnotesize{}{}-14.3 (5.4)}  &  & {\footnotesize{}{}-8.16 (32.7)}\tabularnewline
{\footnotesize{}{}Household Size}  &  & {\footnotesize{}{}5.6 (2.1)}  &  & {\footnotesize{}{}3.3 (18.8)}\tabularnewline
{\footnotesize{}{}Head Age}  &  & {\footnotesize{}{}-1.0 (0.6)}  &  & {\footnotesize{}{}-0.53 (17.0)}\tabularnewline
{\footnotesize{}{}Education}  &  & {\footnotesize{}{}-39.9 (21.5)}  &  & {\footnotesize{}{}-19.2 (34.2)}\\[.7em]
 &  & \multicolumn{3}{c}{\textbf{\footnotesize{}{}Impact Weights $\mathbf{\lambda}$}}\tabularnewline
{\footnotesize{}{}Missed Schooling (per day)}  &  & {\footnotesize{}{}-0.03 (0.17)}  &  & {\footnotesize{}{}0.02 (1.26)}\tabularnewline
 \footnotesize{Sickness (per child sick day)}  &  & {\footnotesize{}{}0.08 (0.05)}  &  & {\footnotesize{}{}0.11 (0.07)}\tabularnewline
 \footnotesize{Value Regardless of Impact}  &  & {\footnotesize{}{}0.47 (3.75)}  &  & {\footnotesize{}{}0.75 (109.21)}\tabularnewline
$\sigma$  &  & {\footnotesize{}{}0.17 (0.17)}  &  & {\footnotesize{}{}0.10 (0.1)}\tabularnewline
{\footnotesize{}{}N}  &  & {\footnotesize{}{}6642}  &  & {\footnotesize{}{}6642}\tabularnewline
\bottomrule
\end{tabular}
\begin{tablenotes}[normal,flushleft]
\footnotesize
\item \emph{Notes}: Both columns computed using our method, using heterogeneous treatment effects estimated with causal forest (see Figure~\ref{fig:Distribution-of-Estimated-CF}). Standard errors are computed using a two-step bootstrap procedure that accounts for uncertainty in both treatment effects and preference parameters. 
% Observations are drawn with replacement before estimation of the treatment effects and the welfare and impact weights. Treatment effects are then estimated from these bootstrapped samples, and welfare and impact weights estimated from these bootstrapped treatment effect estimates; the standard errors reported are the standard deviation across bootstrapped welfare and impact weight estimates. 
We exclude bootstrap draws (0 draws for full ranking, 1 draw for binary ranking, out of 50 for each) that converged to corner solutions against the zero lower bound for omega.
\end{tablenotes}
\end{threeparttable}
\end{table} 
%%%%%%%%%%%%%%%%%%%%%%%%%%%%%%%%%%%%%%%%%%%%%%%

%%%%%%%%%%%%% TABLE %%%%%%%%%%%%%%%%%%%%%%%
\begin{table}
\centering \caption{Treatment Effect Coefficient Estimates: OLS\label{tab:Treatment-Effect-Coefficient-OLS}}
\begin{threeparttable} %
\begin{tabular}{l>{\raggedright}p{3cm}>{\raggedright}p{3cm}>{\raggedright}p{3cm}}
 \toprule
 & {\footnotesize{}{}Log Consumption }  & {\footnotesize{}{}Schooling} & {\footnotesize{}{}Health }
 \tabularnewline
 & {\footnotesize{}{}(Monthly avg. per person, in pesos) }  & {\footnotesize{}{}(Avg. days school missed per child)} & {\footnotesize{}{}(Avg. sick days per child) } \tabularnewline
\midrule 
{\footnotesize{}{}Treatment }  & {\footnotesize{}{}-0.0271 (0.115)}  & {\footnotesize{}{}-0.3527 (0.249)} & {\footnotesize{}{}-0.8111 (0.456)} \tabularnewline
{\footnotesize{}{}Treatment X head indigenous }  & {\footnotesize{}{}0.1462 (0.027)}  & {\footnotesize{}{}0.0323 (0.058)} & {\footnotesize{}{}0.0389 (0.106)} \tabularnewline
{\footnotesize{}{}Treatment X log(Income 1997) }  & {\footnotesize{}{}-0.0027 (0.02)}  & {\footnotesize{}{}0.0141 (0.042)} & {\footnotesize{}{}0.06 (0.077)} \tabularnewline
{\footnotesize{}{}Treatment X num adults}  & {\footnotesize{}{}-0.0157 (0.012)}  & {\footnotesize{}{}0.0134 (0.027)} & {\footnotesize{}{}0.0086 (0.05)} \tabularnewline
{\footnotesize{}{}Treatment X head age }  & {\footnotesize{}{}0.0017 (0.002)}  & {\footnotesize{}{}0.0067 (0.004)} & {\footnotesize{}{}0.0046 (0.007)} \tabularnewline
{\footnotesize{}{}Treatment X head education }  & {\footnotesize{}{}0.0208 (0.003)}  & {\footnotesize{}{}-0.0102 (0.007)} & {\footnotesize{}{}0.0121 (0.014)} \tabularnewline
{\footnotesize{}{}Treatment X male head of household }  & {\footnotesize{}{}-0.0379 (0.063)}  & {\footnotesize{}{}0.1006 (0.137)} & {\footnotesize{}{}0.1513 (0.251)} \tabularnewline
{\footnotesize{}{}Treatment X head agricultural worker }  & {\footnotesize{}{}0.0523 (0.029)}  & {\footnotesize{}{}-0.11 (0.062)} & {\footnotesize{}{}-0.0803 (0.114)} \tabularnewline
{\footnotesize{}{}Treatment X num child less than 2 yrs }  & {\footnotesize{}{}-0.0297 (0.016)}  & {\footnotesize{}{}0.0907 (0.034)} & {\footnotesize{}{}0.0019 (0.063)} \tabularnewline
{\footnotesize{}{}Treatment X num child 3 to 5 yrs }  & {\footnotesize{}{}-0.0307 (0.019)}  & {\footnotesize{}{}-0.0604 (0.041)} & {\footnotesize{}{}0.1187 (0.075)} \tabularnewline
{\footnotesize{}{}Treatment X num child 6 to 10 yrs }  & {\footnotesize{}{}0.0451 (0.015)}  & {\footnotesize{}{}0.015 (0.032)} & {\footnotesize{}{}-0.0178 (0.058)} \tabularnewline
{\footnotesize{}{}Treatment X num boys 11 to 14 yrs }  & {\footnotesize{}{}0.0139 (0.021)}  & {\footnotesize{}{}-0.0495 (0.045)} & {\footnotesize{}{}-0.0218 (0.082)} \tabularnewline
{\footnotesize{}{}Treatment X num girls 11 to 14 yrs }  & {\footnotesize{}{}0.0212 (0.021)}  & {\footnotesize{}{}-0.0279 (0.045)} & {\footnotesize{}{}0.0315 (0.082)} \tabularnewline
{\footnotesize{}{}Treatment X num boys 15 to 19 yrs }  & {\footnotesize{}{}-0.024 (0.023)}  & {\footnotesize{}{}0.001 (0.051)} & {\footnotesize{}{}-0.0633 (0.093)} \tabularnewline
{\footnotesize{}{}Treatment X num girls 15 to 19 yrs }  & {\footnotesize{}{}-0.0243 (0.023)}  & {\footnotesize{}{}0.0035 (0.049)} & {\footnotesize{}{}0.0558 (0.09)} \tabularnewline
{\footnotesize{}{}Treatment X num men 20 to 34 yrs }  & {\footnotesize{}{}0.0037 (0.027)}  & {\footnotesize{}{}-0.0224 (0.058)} & {\footnotesize{}{}-0.2424 (0.106)} \tabularnewline
{\footnotesize{}{}Treatment X num women 20 to 34 yrs }  & {\footnotesize{}{}0.0044 (0.028)}  & {\footnotesize{}{}0.0439 (0.062)} & {\footnotesize{}{}0.1897 (0.113)} \tabularnewline
{\footnotesize{}{}Treatment X num men 35 to 54 yrs }  & {\footnotesize{}{}0.0479 (0.039)}  & {\footnotesize{}{}-0.0492 (0.084)} & {\footnotesize{}{}0.0115 (0.153)} \tabularnewline
{\footnotesize{}{}Treatment X num women 35 to 54 yrs }  & {\footnotesize{}{}-0.0225 (0.037)}  & {\footnotesize{}{}0.0198 (0.079)} & {\footnotesize{}{}-0.0142 (0.145)} \tabularnewline
{\footnotesize{}{}Treatment X num men at least 55 yrs }  & {\footnotesize{}{}-0.0416 (0.053)}  & {\footnotesize{}{}-0.3011 (0.116)} & {\footnotesize{}{}-0.0174 (0.212)} \tabularnewline
{\footnotesize{}{}Treatment X num women at least 55 yrs }  & {\footnotesize{}{}-0.0283 (0.041)}  & {\footnotesize{}{}0.0372 (0.089)} & {\footnotesize{}{}-0.0739 (0.163)} \tabularnewline
\midrule 
{\footnotesize{}{}Baseline Covariates }  & {\footnotesize{}{}X }  & {\footnotesize{}{}X }  & {\footnotesize{}{}X } \tabularnewline
{\footnotesize{}{}$R^{2}$ }  & {\footnotesize{}{}0.180}  & {\footnotesize{}{}0.0132} & {\footnotesize{}{}0.03} \tabularnewline
{\footnotesize{}{}N }  & {\footnotesize{}{}6642 }  & {\footnotesize{}{}6642} & {\footnotesize{}{}6642 } \tabularnewline
\bottomrule
\end{tabular}\begin{tablenotes}[normal,flushleft]
\item \footnotesize \emph{Notes:} OLS coefficients of household characteristics interacted with treatment effects on  three outcome dimensions: log consumption (log monthly per capita consumption), schooling (number of missed school days per child), and health (number of sick days per child). Schooling and health sick days / missed school days measured over 28 days prior to survey. Baseline covariates here includes the covariates without interaction
with treatment effects, e.g. head age, as well as a constant term.
\end{tablenotes}
\end{threeparttable} 
\end{table}

%%%%%%%%%%%%%%%%%%%%%%%%%%%%%%%%%%%%%%

%\centering{\Large{\textbf{Appendix Figures}}}

\begin{landscape}
\begin{figure}[b]
\caption{Binscatter Plots of Treatment Effect Heterogeneity: Causal Forest \label{fig:Binscatter-CF}}
\begin{subfigure}[t]{.5\textwidth}
\includegraphics[width=\columnwidth]{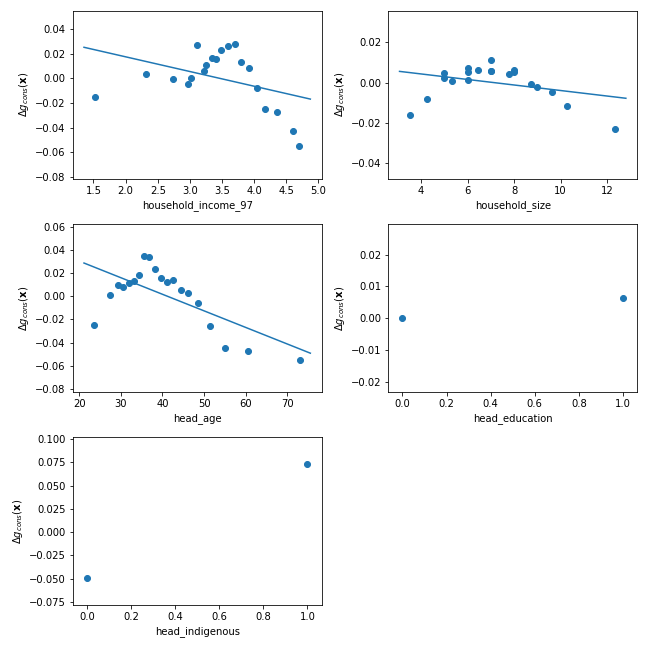}
\caption{Log Consumption Treatment Effects}
\end{subfigure}
\begin{subfigure}[t]{.5\textwidth}
\includegraphics[width=\columnwidth]{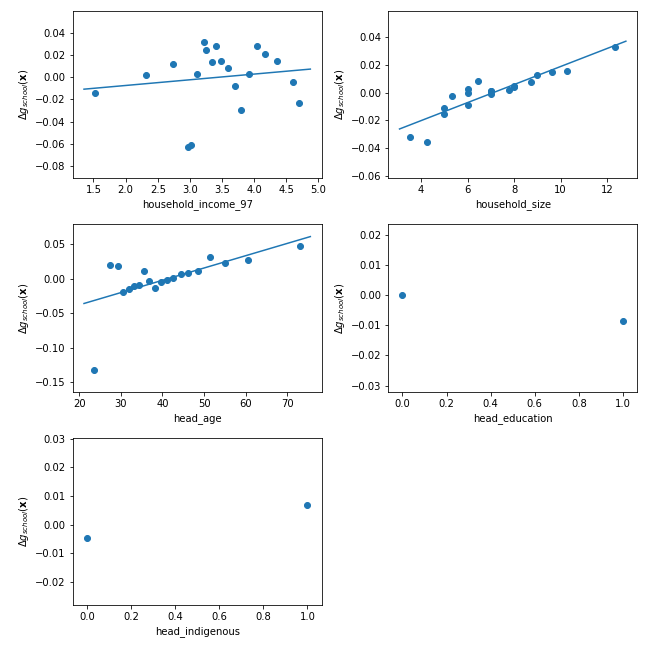}
\caption{Schooling Treatment Effects}
\end{subfigure}
\begin{subfigure}[t]{.5\textwidth}
\includegraphics[width=\columnwidth]{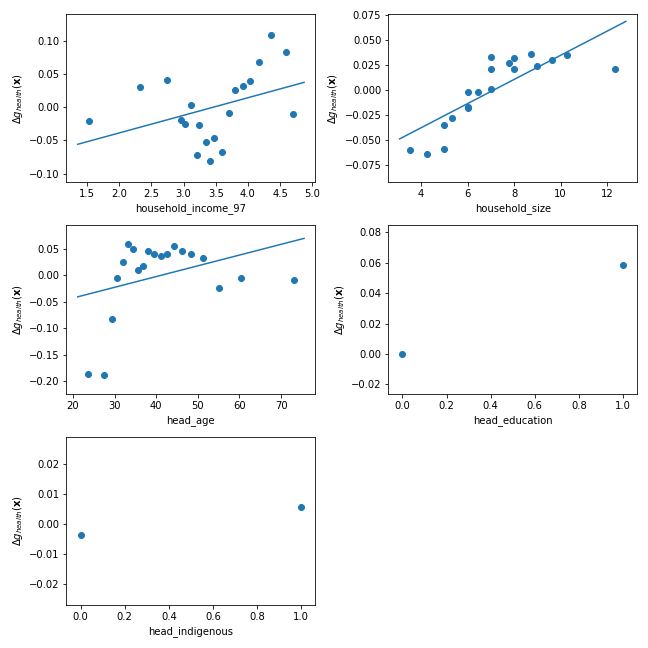}
\caption{Health Treatment Effects}
\vspace{.3cm}
\end{subfigure}
{\footnotesize \emph{Notes}: Binscatter plots of treatment effects from causal forest over a selected group of five covariates: household size; household head education; household head indigenous status; household head age;
and log household income in the pre-period of 1997. Figures shown for treatment effects over per-person monthly consumption, number of sick days per child, and number of missed school days per child. Treatment effects shown are residualized against remaining covariates in the regression (the other graphed covariates).}
\end{figure}
\end{landscape}

\begin{landscape}
\begin{figure}[b]
\caption{Binscatter Plots of Treatment Effect Heterogeneity: OLS \label{fig:Binscatter-OLS}}
\begin{subfigure}[t]{.5\textwidth}
\includegraphics[width=\columnwidth]{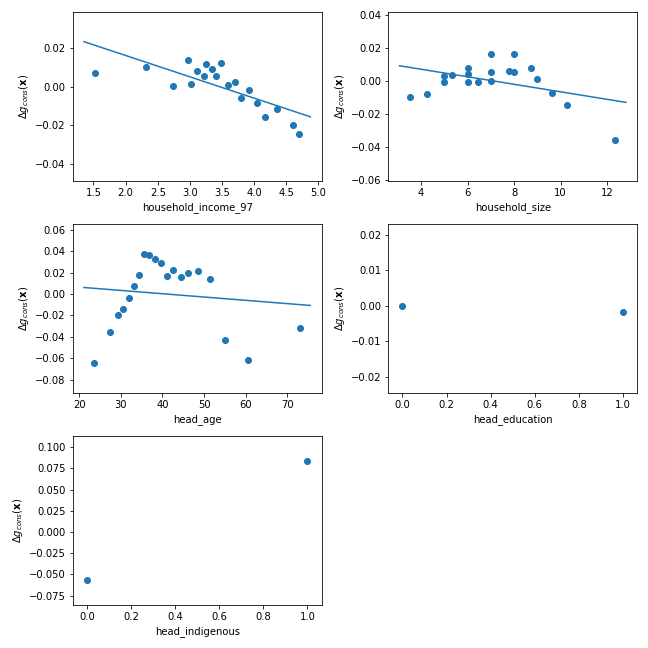}
\caption{Log Consumption Treatment Effects}
\end{subfigure}
\begin{subfigure}[t]{.5\textwidth}
\includegraphics[width=\columnwidth]{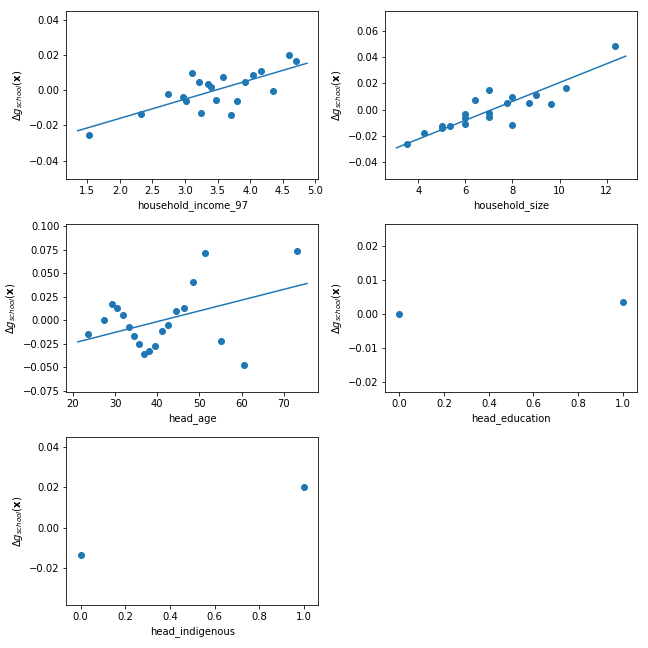}
\caption{Schooling Treatment Effects}
\end{subfigure}
\begin{subfigure}[t]{.5\textwidth}
\includegraphics[width=\columnwidth]{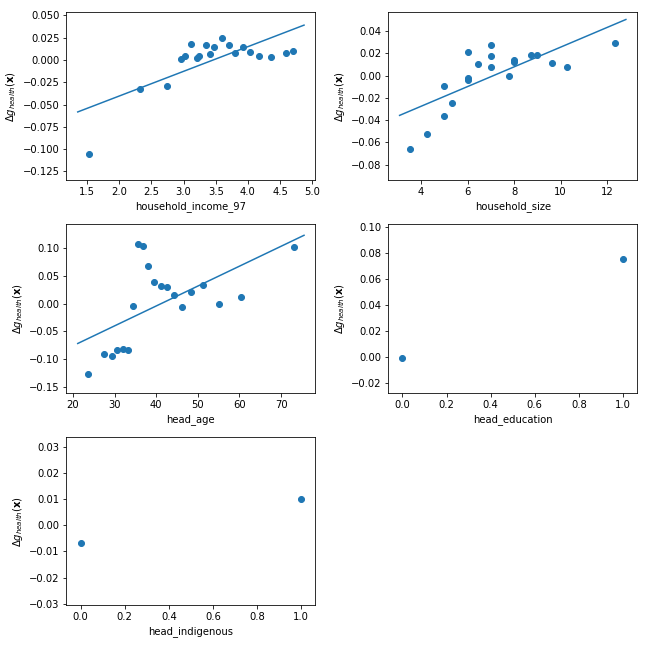}
\caption{Health Treatment Effects}
\vspace{.3cm}
\end{subfigure}
{\footnotesize \emph{Notes}:  Binscatter plots of treatment effects from OLS over five covariates: household size; household head education; household head indigenous status; household head age; and log household income in the pre-period of 1997. Figures shown for treatment effects over per-person monthly consumption, number of sick days per child, and number of missed school days per child.
Treatment effects shown are residualized against remaining covariates in the regression (the other graphed covariates).}
\end{figure}
\end{landscape}

\begin{figure}
\begin{centering}
\caption{Expected Program Impacts under Alternative Preferences, in Welfare Space}\label{fig:allocations3d_survey}

\begin{subfigure}[tc]{\textwidth}
    \centering
    \includegraphics[width=0.6\columnwidth]{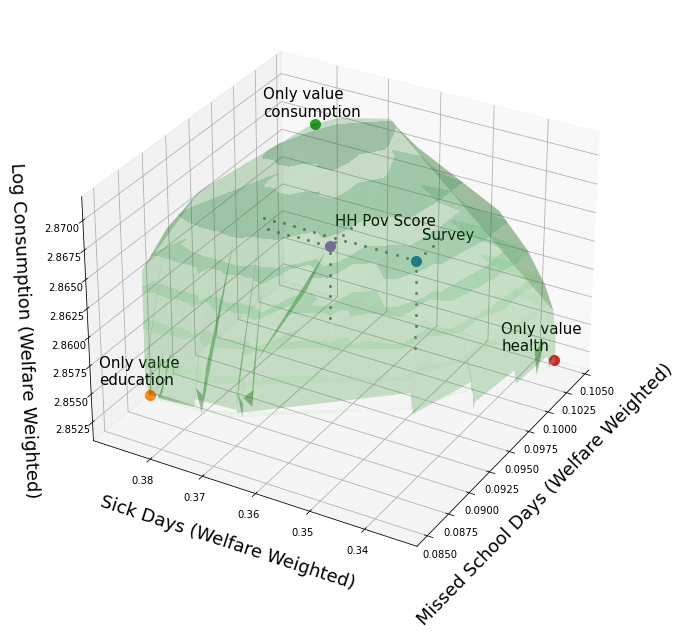}
    \caption{Welfare-Weighted Outcomes}
\end{subfigure}
\begin{subfigure}[tc]{\textwidth}
    \centering
    \includegraphics[width=0.6\columnwidth]{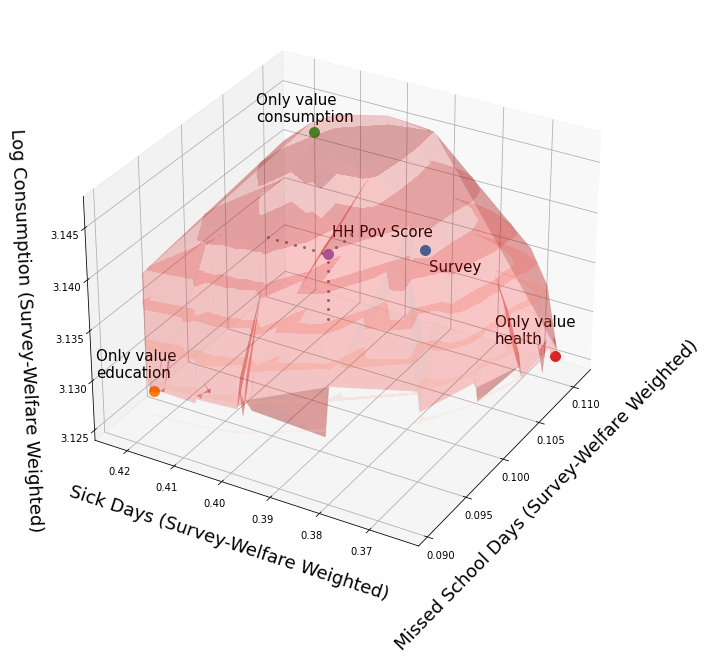}
    \caption{Survey Welfare-Weighted Outcomes}
\end{subfigure}

\end{centering}
    {\footnotesize \textit{Notes:} Figure shows the frontier of outcomes resulting from all possible allocations, weighted by welfare weights. Panel (a) weights outcomes by the welfare weights consistent with the implemented PROGRESA program, derived using our method. Panel (b) uses welfare weights consistent with the inferred preferences from the implemented survey. Labeled points correspond to particular allocations described in Table \ref{tab:Allocation-Rules}.}
\end{figure}

\end{document}